\documentclass[prb,twocolumn,showpacs,superscriptaddress]{revtex4-1}

\usepackage{amsmath, amssymb}    
\usepackage{graphicx}   
\usepackage{subfigure}
\usepackage{verbatim}   
\usepackage{color}      
\usepackage{subfigure}  
\usepackage{hyperref}   

\renewcommand{\v}[1]{{\bf #1}}
\newcommand{\Eq}[1]{Eq.~(\ref{#1})}
\newcommand{\Fig}[1]{Fig.~\ref{#1}}

\newcommand{\cL}[0]{\ensuremath{\cal L}}
\newcommand{\cH}[0]{\ensuremath{\cal H}}
\newcommand{\cN}[0]{{\cal N}}
\newcommand{\cS}[0]{{\cal S}}

\def\bd{\begin{displaymath}}
\def\be{\begin{equation}}
\def\ed{\end{displaymath}}
\def\ee{\end{equation}}
\def\bsub{\begin{subequations}}
\def\esub{\end{subequations}}

\begin{document}


\title{Heat equation approach to geometric changes of the torus Laughlin-state}

\author{Zhenyu Zhou}
\affiliation{Department of Physics and Center for Materials Innovation, Washington University, St. Louis, MO 63130, USA}
\author{Zohar Nussinov}
\affiliation{Department of Physics and Center for Materials Innovation, Washington University, St. Louis, MO 63130, USA}
\author{Alexander Seidel}
\affiliation{Department of Physics and Center for Materials Innovation, Washington University, St. Louis, MO 63130, USA}


\date{\today}

\begin{abstract}

We study the second quantized -or guiding center- description of the torus Laughlin state.
Our main focus is the change of the guiding center degrees of freedom with the torus geometry,
which we show to be generated by a two-body operator. 
We demonstrate that this operator can be used to evolve the full torus Laughlin state
at given modular parameter $\tau$ from its simple (Slater-determinant) thin torus limit,
thus giving rise to a new presentation of the torus Laughlin state in terms of its
``root partition'' and an exponential of a two-body operator.
This operator therefore generates in particular the adiabatic evolution between
Laughlin states on regular tori and the quasi-one-dimensional thin torus limit.
We make contact with the recently introduced notion of a ``Hall viscosity'' for 
fractional quantum Hall states, to which our two-body operator is naturally related,
and which serves as a demonstration of our method to generate the Laughlin state on the torus.

\end{abstract}

\pacs{PACS}

\maketitle

\section{Introduction}

The discovery of the fractional quantum Hall effect\cite{stormer82} 
has led to a series of remarkable theoretical developments.
Much insight has flowed from principles that govern the construction
of certain ``special wave functions''\cite{laughlin, MR, RR, haldane_rezayi88, halperin83,  gaffnian} and their relation
with conformal field theory (CFT)\cite{MR}, and/or their interpretation in a composite fermion picture\cite{composite_fermion}.
The credibility of this approach is greatly enhanced by the construction of parent Hamiltonians
for such wave functions, which has been possible in many interesting cases.\cite{RR, haldane_rezayi88, haldane_hierarchy, trugman85, greiter91}  
This very particular class of quasi-solvable\footnote{By this we mean that the ground state is exactly known.} Hamiltonians
consists of Landau level projected ultra-local interactions, which enforce the analytic properties that uniquely characterize
the respective ground state. The prime example for such a parent Hamiltonian is given by the $V_1$-pseudopotential,\cite{haldane_hierarchy} which is a pairwise (two-particle) projection operator onto states of relative angular momentum $1$
within the lowest Landau level (LLL).
Its unique ground state at filling factor $\nu=1/3$ is the Laughlin state corresponding to this filling.

Due to the Landau level projection, the pseudo-potential Hamiltonian acts only on the ``guiding center''
degrees of freedom, which exhaust the large degeneracy within a given Landau level, and commute with the generators
of inter-level transitions. (The latter are related to the kinetic momenta of the particles, see below. For a review of physics in a magnetic
field, the reader is referred to Ref. \onlinecite{macdonald94}.)
It is therefore beneficial to make the action of the Hamiltonian on guiding center variables manifest.
This is in particular the case when the Hamiltonian is expressed
using creation/annihilation  operators for a set of eigenstates, say,
of one of the two non-commuting guiding center
components, which form a basis for the LLL. 
In numerics,  such a second quantized ``guiding center'' description of the Hamiltonian
is essential to make use of the
reduced Hilbert space dimensionality owing to the LLL projection. 
We illustrate this procedure for the cylinder geometry, for reasons that will soon become apparent. To this end,
we introduce a set of LLL basis states as described above, given by:
\begin{equation}\label{phi_cyl}
\phi_n(z) = \xi^n e^{-\frac{1}{2} x^2} e^{-\frac{1}{2} n^2/r^2}\,
\end{equation}
where $\xi=e^{z/r}$ is an analytic function of $z=x+iy$ that satisfies 
periodic boundary conditions in $y$, appropriate for a cylinder of perimeter $2\pi r$
(using Landau gauge, $\v A=(0,x)$).
These orbitals are eigenstates of the x-component of the guiding center
with eigenvalues $n/r$, where, for the time being, we set the magnetic length $l_B$
equal to $1$. The 1/3-Laughlin state on the cylinder is then expressed as\cite{rezayi_haldane94}
\begin{equation}\label{laugh_cyl}
\psi_{1/3}(z_1\dotsc z_N)=\prod_{i<j} (\xi_i-\xi_j)^3 \times e^{-\frac{1}{2} \sum_k x_k^2}\,.
\end{equation}

With respect to  the basis \Eq{phi_cyl}, the $V_1$ pseudo-potential 
takes on the following second quantized form (cf, e.g., Ref. \onlinecite{Lee_Leinaas04}):
\begin{equation}
\begin{split}\label{V1_cyl}
\hat V_1&=\sum_R Q^\dagger_R Q_R\\
 Q_R&=\sum_x x \exp(-x^2/r^2) \,c_{R-x} c_{R+x}
\end{split}
\end{equation}
In the first line, the sum goes over both integer and half-odd integer values of $R$,
whereas in the second it goes over integer (half-odd integer) if $R$ is integer (half-odd integer),
such that labels $R\pm x$ are then always integer.

The one parameter family of models \eqref{V1_cyl} share many features with one-dimensional (1D)
lattice
models that arise elsewhere in solid state physics, such as translational invariance and short ranged
(exponentially decaying) interactions. It is thus not surprising that it has recently been proposed 
to be of use in the absence of (proper) Landau level physics, e.g., in flat band solids both
with\cite{qi11} and without\cite{scarola11} non-zero Chern numbers, and in quite general terms in
Ref. \onlinecite{seidel1}.

Despite the usefulness of the second quantized description \eqref{V1_cyl} of the pseudo-potential, it would be very difficult
to solve for the zero energy eigenstates of the model in this form, or to even know analytically that such
zero energy eigenstates exist. For this we rely on the original first-quantized definition of the pseudo-potential $\hat V_1$,
and on the explicitly known analytic form 
of the Laughlin state, \Eq{laugh_cyl}, in terms of ordinary position variables. 
It would be highly non-trivial, however, to come up with such a first
quantized language for the problem \Eq{V1_cyl} if its connection to LLL orbitals were not a priori known.
This is so because this language becomes available only after proper embedding of the degrees of freedom
associated with the operators $c_n$, $c_n^\dagger$ in \Eq{V1_cyl} into a larger Hilbert space. In \Eq{V1_cyl}, no information
is retained about the kinetic momenta that determine the structure of the Landau levels.
Indeed, as Haldane has recently shown,\cite{haldane11}
by making these kinetic momenta subject to a different metric from that entering the interactions,
one obtains a different way to naturally embed the problem \eqref{V1_cyl}
into the larger Hilbert space of square integrable functions. In this setup, \Eq{V1_cyl} remains unaltered,
but the resulting wave function loses the analytic properties of \Eq{laugh_cyl} that make the problem
tractable.\cite{haldane11,kunyang12} Moreover, the solid state applications mentioned initially represent yet 
another way to embed the problem \eqref{V1_cyl} into a larger Hilbert space.

These considerations show that ``interaction only'' models such as \eqref{V1_cyl},
especially ones that share the ``center-of-mass conserving'' property,\cite{seidel1}
may enjoy a considerable range of applications, but at the same time, may  be quite hard to solve in general.\footnote{We note though
a tractable truncated version of \Eq{V1_cyl} with matrix product ground state given in Ref. \onlinecite{Bergholtz12}.}
This is chiefly due to the fact that the Laughlin state, in its second quantized/guiding center presentation,
is quite a bit more complicated than in its analytic first quantized form \Eq{laugh_cyl}.
While no closed form seems to be known for the amplitudes $\langle 0|c_{n_1}\dotsc c_{n_N}|\psi_{1/3}\rangle$,
much progress has recently been made in understanding their structure for the cylinder geometry,
and for any other geometry in which the analytic part of Laughlin's wave function is given by a polynomial.
Indeed, for Laughlin states and many other quantum Hall trial wave functions, these polynomials have been identified as
Jack polynomials, multiplied by Jastrow factors.\cite{bernevig08, bernevig08-2}
This allows the amplitudes $\langle 0|c_{n_1}\dotsc c_{n_N}|\psi_{1/3}\rangle$ to be determined recursively.
For the cylinder Laughlin state, this can be sketched as follows. We consider the expansion of \Eq{laugh_cyl} into monomials,
\begin{equation}
\begin{split}\label{V1_cyl2}
\psi_{1/3}(z_1\dotsc z_N)=\sum_{\{n_k\}} C_{\{n_k\}}\prod_k \xi_k^{n_k} e^{-\frac{1}{2} x_k^2}
\end{split}
\end{equation}
The product in the above equation can be interpreted as a state with definite single particle occupation numbers,
up to a normalization. (The $C_{\{n_k\}}$ have the proper (anti)-symmetry to allow (anti)-symmetrization of the product.)
This normalization is readily read from \Eq{phi_cyl}. We thus have\cite{rezayi_haldane94}
\begin{equation}\label{ket_cyl}
|\psi_{1/3}\rangle_r = \sum _{\{n_k\}}   e^{\frac{1}{2r^2} \sum_k n_k^2} C_{\{n_k\}} \;c^\dagger_{n_N}\dotsc c^\dagger_{n_1} |0\rangle\,.
\end{equation}
The monomial coefficients do not depend on $r$, and are known recursively, starting from the coefficient of the ``root configuration''
$c^\dagger_{n_N}\dotsc c^\dagger_{n_1} |0\rangle=|10010010010\dotsc\rangle$ through a process known as ``inward squeezing''.\cite{bernevig08, bernevig08-2}

A remarkable aspect of \Eq{ket_cyl} is that the dependence on geometry, in this case the cylinder radius $r$, comes in only through
the trivial normalization factor. This is matched by a similarly trivial $r$-dependence of the interaction $\hat V_1$.
It is quite easy to see that the condition that the Hamiltonian \Eq{V1_cyl} has a zero energy eigenstate (which, by positive semi-definiteness, must be a ground state), which reduces to $Q_R|\psi_{1/3}\rangle_r=0$ $\forall R$, yields an $r$-independent condition on the coefficients $C_{\{n_k\}}$. In this way it becomes manifest that regardless of the value or $r$, one is always solving the same problem,
which is intuitively clear from the simple analytic form of the Laughlin wave function \eqref{laugh_cyl} and its trivial $r$-dependence.
It should also be emphasized that the simple $r$-dependence of \Eq{ket_cyl} is not particular to the Laughlin state. It is a direct consequence of the polynomial form of the wave function, and carries over without change to any quantum Hall trial state on the cylinder.

The situation is rather different for the torus geometry. 
The main purpose of this work will be to get a handle on the guiding center presentation of the torus Laughlin states.
In the remainder of this introduction, we review some well known facts that make life
more complicated on the torus.

\begin{figure}[t!]
  \centering
    \includegraphics[width=0.45\textwidth]{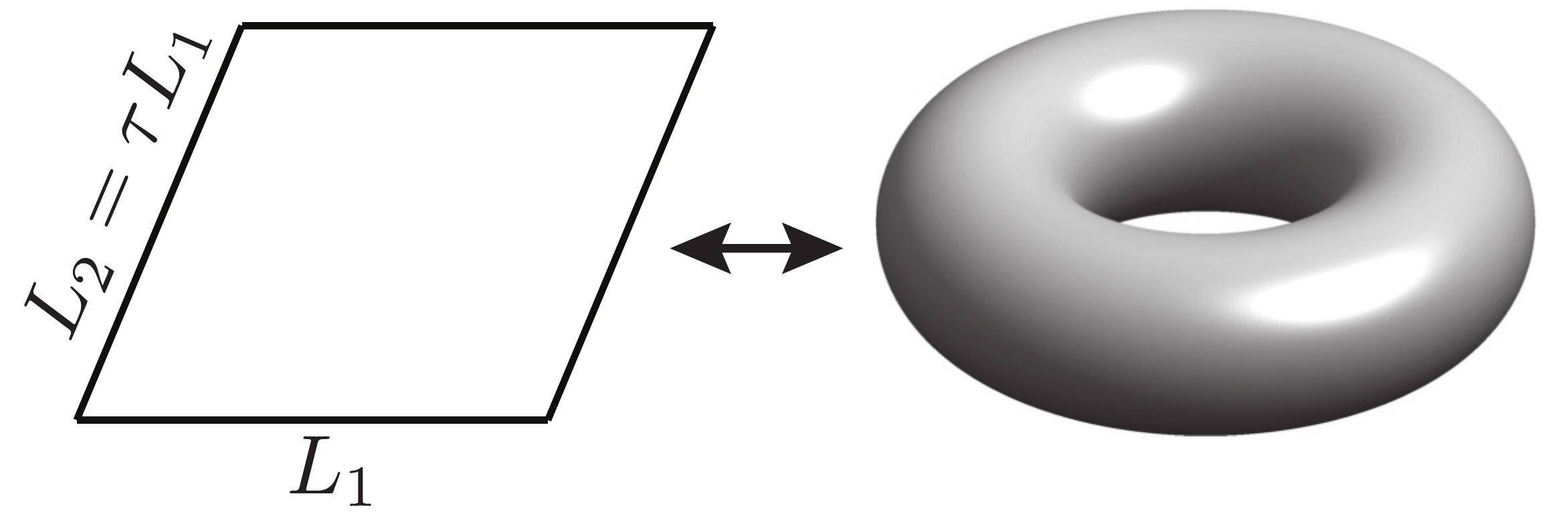}
     \caption{\label{domain} Fundamental domain for torus wave functions.}
\end{figure}

In first quantized language, we pass to the torus by introducing
periodic boundary conditions in the complex plane along two  fundamental periods $L_1$ and $L_2$, where $L_1$ is taken to
be real, and $\mbox{Im} L_2>0$ (\Fig{domain}). The geometry of the torus can be parameterized by $\tau=L_2/L_1$, the modular parameter. The Laughlin state at general filling factor $1/q$ then becomes\cite{haldane85}
\begin{equation}\label{laugh_tor}
\psi_{1/q}^\ell(z_1\dotsc z_N)=\exp(-\frac{1}{2}\sum_k y_k^2)\,F^{\ell}(Z,\tau) \prod_{i<j} \theta_1 (\frac{z_i-z_j}{L_1},\tau)^q \,
\end{equation}
Here, $\theta_1(z,\tau)$ is the odd Jacobi theta-function, and for the factor depending on the ``center of mass'' $Z=z_1+\dotsc z_N$,
which also depends on an additional label $\ell=0\dotsc q-1$ corresponding to a choice of basis in the $q$-fold degenerate\cite{wen_niu90}
ground state space, we adopt the convention of Ref. \onlinecite{read_rezayi96}:
\begin{equation}\label{CM}
F^\ell(Z,\tau)=\theta\left[ \begin{array}{c} \frac{\ell}{q}  +\frac{L-q}{2q}\\ -\frac{L-q}{2}\end{array}\right](qZ/L_1, q\tau)\,.
\end{equation}
Here, $\theta{a \brack b}(z,\tau)$ is the Jacobi theta function of characteristics $a$ and $b$, and $L=L_1 \mbox{Im} L_2/(2\pi) $
is the number of flux quanta penetrating the surface of the torus.

Thus, while the Laughlin state is still of the general form of a Gaussian factor multiplying an analytic function in the
complex particle coordinates $z_i$, the latter is not of polynomial form. As a result, to the best of our knowledge,
there is currently no detailed understanding of the structure of the guiding center description of this state.
By this we mean a general understanding of the coefficients of the analog of
\Eq{ket_cyl}:
\begin{equation}\label{ket_tor}
|\psi_{1/3}^\ell\rangle_\tau = \sum _{\{n_k\}}    C_{\{n_k\}}^\ell(\tau) \;c^\dagger_{n_N}\dotsc c^\dagger_{n_1} |0\rangle\,.
\end{equation}
In particular, the $\tau$-dependence of the coefficients $C^\ell_{\{n_k\}}(\tau)$ is {\em not} of a simple form 
reminiscent of the $r$-dependence explicit in \Eq{ket_cyl}.
Moreover, intuitively, one would still expect that these coefficients can be generated from the dominance pattern, i.e.,
$100100100\dotsc$ at $\nu=1/3$. Indeed, this configuration is still dominant on the torus, in the sense that it is the configuration
that dominates in the thin torus limit\cite{seidel1,karlhede1,seidel_dunghai06,karlhede2}.
 The success of the thin torus approach in determining physical properties,
such as Abelian and non-Abelian statistics\cite{ seidel_dunghai07, seidel2008a, john_seidel11, john_seidel12} 
and the presence of gapless excitations\cite{seidel11}, suggests that
even on the torus these patterns allow for a reconstruction of the full many-body wave function.
On the other hand, there is no notion of ``inward'' squeezing on the torus, due to periodic boundary conditions.
The main result of this paper will be the development of a machinery for the above mentioned reconstruction of the full
torus Laughlin state in the guiding center description, from the thin torus state. Since as an additional complication, such machinery can be expected to depend
non-trivially on $\tau$, we first focus our attention on the dependence of the coefficients $C^\ell_{\{n_k\}}(\tau)$ on the geometric parameter.

As a final remark, we point out\cite{haldane85} that the torus Laughlin states at $\nu=1/3$  are still the unique ground state of the $V_1$ pseudo-potential.
Its second quantized form agrees with a straightforward periodization of the model \eqref{V1_cyl}, with
\begin{equation}\label{V1_tor}
 Q_R=\sum_{0<x<L/2 \atop x+R\in \mathbb{Z}} \;\sum_{m \in \mathbb{Z}} (x+mL) \exp[\frac{2\pi i \tau}{L}(x+mL)^2] \,c_{R-x} c_{R+x}\,.
\end{equation}
One sees that for $\tau= i |L_2|/ 2\pi  r$, $L_2=iL/r$, this reduces to the cylinder form \Eq{V1_cyl} for $L\rightarrow\infty$,
and respects the periodic boundary condition $c_n=c_{n+L}$ otherwise. (\Eq{V1_tor} is valid for general complex $\tau$, though).
One therefore passes from \Eq{V1_cyl} to \Eq{V1_tor} (with imaginary $\tau$) through straightforward introduction of periodic
boundary conditions (PBCs). Yet the solution of \Eq{V1_tor} is arguably much less under control.
The introduction of PBCs is a standard and very useful tool throughout solid state physics. We thus expect that 
a better understanding of the guiding center description of the torus Laughlin state will also benefit
the solid state applications\cite{scarola11, qi11} mentioned initially.

The remainder of the paper is organized as follows.
In section \ref{Gen} we construct a two-body operators that generated the
changes in the guiding center variables of the torus Laughlin state with modular
parameter $\tau$. Sections \ref{cylinder} and \ref{torus_intro} highlight further formal similarities and
differences between the cylinder and the torus.
Sec. \ref{heatequation} presents the heat equation for the $\tau$-derivative
of the analytic Laughlin state. Sec. \ref{mapping} introduces a 2D to 1D mapping, which is
our device for embedding lowest Landau levels at different modular parameter $\tau$
into the same larger Hilbert space. In Sec. \ref{def_Gt} we derive the generator
mentioned above. In Sec. \ref{symmetries} we symmetrize this operator and present a byproduct of this
study, a hitherto unknown class of two-body operators that annihilate the torus Laughlin state.
In Sec. \ref{present} we postulate a presentation of the torus Laughlin state in terms of its
thin torus, or ``dominance'' pattern, and the class of two-body operators generating
changes in geometry. In Sec. \ref{viscosity} we demonstrate the postulate of Sec. \ref{present}
numerically, and work out the relation of our generator with the Hall viscosity\cite{read09}, 
 which we calculate numerically as a demonstration of analytical results,
comparing the resulting data to earlier numerical studies.
We discuss our results in Sec. \ref{discussion}, and conclude in Sec. \ref{conclusion}.
A small Appendix discusses a minor technical detail.

\section{Construction of the 2-body operator\label{Gen}}

\subsection{A final look at the cylinder case\label{cylinder}}

As motivated above, to establish a machinery that generates the full guiding center description
of the Laughlin state from the root configuration, a natural starting point is to get under control
how this description changes with the geometric parameter $\tau$. To this end, we will seek to 
construct an operator that generates changes of the guiding center degrees of freedom to first
order in $d\tau$. The similar problem for the cylinder, where $r$ is the geometric parameter, is
comparatively trivial and was
already addressed in the introduction. For later reference, it is instructive to first cast these results
in terms of a generator of infinitesimal changes in the parameter $r^{-2}$.
\Eq{ket_cyl} can be written as
\be\label{exp_cyl}
   |\psi_{1/3}\rangle_{r'} = e^{(r'^{-2}-r^{-2}) G_{r^{-2}}}  |\psi_{1/3}\rangle_r
\ee
where
\be
G_{r^{-2}}=\frac{1}{2}\sum_{n} n^2 c^\dagger_n c_n\label{Gr}
\ee
is the generator of changes in the geometric parameter $r^{-2}$. Note that it is independent of $r$.
We emphasize again that \eqref{exp_cyl}, \eqref{Gr} are very general, and apply to other quantum Hall trial 
states on the cylinder as well.
In writing \eqref{exp_cyl}, we leave it understood that the exponentiated operator generates the change of the
guiding center degrees of freedom only; it does not generate in any way the change of the LLL orbitals
themselves as a function of $r$, \Eq{phi_cyl}. We are only concerned here with the change in guiding
center degrees of freedom, since the object of study is the second quantized Hamiltonian
\Eq{V1_cyl}, in which degrees of freedom associated with kinetic momenta are not retained.
We will thus carefully distinguish from now on between the Laughlin wave function 
$\psi_{1/3}\equiv \psi_{1/3}(z_1\dotsc z_N;r)$, which lives in the full Hilbert space of square integrable
functions over some domain, and the ket $|\psi_{1/3}\rangle_r$, which lives in an abstract Hilbert space
denoted $\cal L$ that is isomorphic to the LLL for any given value of cylinder radius $r$. Similar conventions
will be used below for the torus. 
In \cL, therefore, all those orbitals with the same guiding center quantum number $n$ become identified,
which originally belonged to different LLLs corresponding to different values of the parameter $r$.
\footnote{Indeed, as formulated at present, these different Landau levels do not even live in
the same Hilbert space, since the domain of the underlying wave functions depends on the value of $r$.
This is inconsequential at present, however, and will later be remedied.}

We note that a similarly universal operator that generates changes of the
guiding center degrees of freedom in response to a change in geometry
can be obtained on the plane.\cite{read_rezayi11, kunyang12}
Here, since a geometric deformation by means of uniform strain does not affect boundary conditions,
such a deformation
is implemented by a change in the metric,
and unlike in \Eq{exp_cyl}, the operation implementing this 
deformation is unitary.

On the other hand,
it is worth pointing out that in \eqref{exp_cyl}, the lack of unitarity leads
to a breakdown of the equation in the ``thin cylinder'' limit $r\rightarrow 0$,
in which $ |\psi_{1/3}\rangle_r$ approaches $|100100100\dotsc\rangle$.
The equation remains valid for arbitrarily small but finite $r$, where the
limiting state $|100100100\dotsc\rangle$ receives arbitrarily small corrections,
which are, however, important and may not be dropped,
since they become large under the non-unitary evolution facilitated by the exponential operator.
This is immediately clear from the fact that the thin cylinder state is an eigenstate of the
one-body operator in the exponent. This operator is thus not capable of generating the off-diagonal matrix element
needed to ``squeeze'' the full many-body wave function out of the thin cylinder state, i.e., the root configuration.
\Eq{exp_cyl} is thus {\em not} a tool to generate the full cylinder Laughlin state out of the root 
configuration. For the cylinder, however, other such tools are already available, as mentioned in the Introduction.\cite{bernevig08, bernevig08-2}

\subsection{General considerations for the generator on the torus\label{torus_intro}}

We desire to construct an operator analogous to $G_{r^{-2}}$ for the torus 
Laughlin state, which generates changes in the guiding center variables
of the state in $\tau$. This operator is thus defined by the following equation:
\be\label{dtau}
\nabla_\tau |\psi_{1/q}^\ell(\tau)\rangle =   {\v G}_\tau |\psi_{1/q}^\ell(\tau)\rangle\,.
\ee
Here, ${\v G}_\tau$ denotes the operator valued two-component object $(G_{\tau_x},G_{\tau_y})$,
and $\nabla_\tau\equiv(\partial_{\tau_x},\partial_{\tau_y})$.
Note that we require that  ${\v G}_\tau$ is {\em independent}
of the label $\ell$ distinguishing the
 $q$ degenerate Laughlin states $|\psi_{1/q}^\ell(\tau)\rangle$,
at given filling factor $1/q$ and given $\tau$.

To highlight considerable differences with the similar problem on the cylinder,
we now show that it follows easily from these assumptions that, unlike for the cylinder,
the components of ${\v G}_\tau$ {\em cannot} be one-body operators.
For, if  $G_{\tau_{x,y}}$ were one-body operators, we could symmetrize each with respect
to the magnetic translation group. After symmetrization, $G_{\tau_{x,y}}$ would still satisfy
\Eq{dtau}. This follows from the observation that  ${\v G}_\tau$ was assumed to be independent of $\ell$,
and that the Laughlin states $|\psi_{1/q}^\ell(\tau)\rangle$ are closed under magnetic translations.
However, the only one-body operator that is invariant under magnetic translations is, up to constants,
the particle number operator $\hat N$. Since the $|\psi_{1/q}^\ell(\tau)\rangle$ are eigenstates of $\hat N$,
it is clear that no such operators could satisfy \Eq{dtau}.

In the following, we will, however, show that $G_{\tau_{x,y}}$ can be a two-body operator.

\subsection{Heat equation for the torus Laughlin state\label{heatequation}}

We begin by deriving a differential equation for the $\tau$ evolution
of the analytic Laughlin wave function \Eq{laugh_tor}. 
We have
\bd
  \partial_\tau \psi_{1/q}^\ell= e^{-\frac{1}{2}\sum_k y_k^2} ((\partial_\tau F^\ell) f_{rel} + F^\ell \partial_\tau f_{rel})
\ed
where $f_{rel}$ denotes the theta-function Jastrow factor in \Eq{laugh_tor}
and $\partial_\tau=\frac{1}{2}(\partial_{\tau_x}-i\partial_{\tau_y})$.
The center-of-mass factor in the form
\Eq{CM} is also given by a theta function. Independent of $\ell$, 
it satisfies the ``heat equation''
\be
\partial_\tau F^\ell (Z,\tau)= \frac{1}{4\pi i q} \partial_Z^2\, F^\ell (Z,\tau)= \frac{1}{4\pi i q} \partial_X^2\, F^\ell (Z,\tau) 
\ee
with $X=\mbox{Re}\, Z$.
Since $\partial_X$ leaves the relative part invariant, the operator $(4\pi i q)^{-1} \partial_X^2$ acting on the torus Laughlin state
produces just the first term above in $\partial_\tau \psi_{1/q}$. The latter can thus be expressed as
\be\label{heat}
\partial_\tau \psi_{1/q}^\ell= \left[\frac{1}{4\pi i q} \partial_X^2 + q\sum_{i<j} \frac{\partial_\tau \theta_1(z_i-z_j,\tau)}{\theta_1(z_i-z_j,\tau)} \right] \psi_{1/q}^\ell\,.
\ee
It is pleasing that the differential operator on the right hand side of the above equation has the form of a two-body 
operator. The are, however, two remaining obstacles before we can express the change of guiding center variables
in terms of a two-body operator derived from the above equation. First, as defined thus far, the Laughlin states
\Eq{laugh_tor} for different parameter $\tau$ do not live in the same Hilbert space. In particular, for fixed $\tau$
the state \eqref{laugh_tor} is usually viewed as a member of the Hilbert space of square integrable functions over
the fundamental domain in \Fig{domain}. In order to view the differential operator in \Eq{heat}
as an operator in some Hilbert space, we must therefore first embed all Laughlin states for different
$\tau$, in fact all the corresponding lowest Landau levels, into the same Hilbert space, since
our differential operator can be viewed as connecting states with infinitesimally {\em different} $\tau$.
The second obstacle is that even with such embedding, the lowest Landau level will depend on $\tau$, i.e.,
 will correspond to a different subspace of the larger Hilbert space $\cal H$ (to be defined below) for different $\tau$.
The differential operator in \Eq{heat} therefore not only describes the change of  guiding center degrees of freedom
with $\tau$, it also describes the change of the Landau level itself, which we are not interested in.
We will therefore find it necessary to extract the piece of \Eq{heat} that acts on guiding centers only.

\subsection{Mapping the problem to 1D\label{mapping}}

We will first address the more technical problem, which is the embedding of the torus Landau levels for 
different $\tau$, denoted by $\cL_\tau$ in the following, into the same larger Hilbert space $\cal H$.
One natural approach that has been emphasized in the recent literature\cite{read09} is to choose an equivalent 
way to formulate the problem, where the fundamental domain remains unchanged and instead the metric 
is deformed. We will return to this point of view in Sec. \ref{viscosity}, where we make connection with the Hall viscosity.

Here we will choose a different approach, which is rooted in the intuition that Landau-level-projected
physics is effectively one-dimensional. One manifestation of this is the form of the ``1D lattice'' Hamiltonian
\Eq{V1_cyl} that governs the guiding center degrees of freedom.
Another is the fact that wave functions in the LLL are entirely determined by holomorphic
functions satisfying certain boundary conditions. As is well known, the values of such functions
in the entire complex plane are already determined by those on (any interval on) the real axis.
For this reason we may restrict our study of the Laughlin states \eqref{laugh_tor}
to the real axis without any loss of information. Also, we find it convenient to
choose $L_1=1$, $L_2=\tau$ as the fundamental domain for the original
two-dimensional (2D) wave functions. With this, after restriction to the real axis,
all states \eqref{laugh_tor} become elements of ${\cH}=L^2[0,1]$ of square
integrable functions within the interval $[0,1]$. We note that with these conventions,
the area of the fundamental domain is not preserved as we change $\tau$.
Therefore, we must accommodate for this by changing the magnetic length
accordingly, such that $\mbox{Im} \tau=2\pi L l_b^2$. This, however,
results only in the following trivial modification of the wave functions \eqref{laugh_tor},
\bd
  \exp(-\frac{1}{2}\sum_k y_k^2)\longrightarrow \exp(-\frac{1}{2}\sum_k y_k^2/l_B^2)\,,
\ed
which is inconsequential since we work at $y=0$ in the following.
Clearly, when \Eq{heat} is now restricted to $y=0$, the operator on the right hand side
is a well defined differential operator within the Hilbert space $\cal H$ (in the usual sense
that its domain is dense in $\cH$.)

A preferred basis for the LLL at given $\tau$, both within the original 2D
as well as the 1D Hilbert space, is given by the following wave functions,
\be\label{chin}
\chi_n(z)=\left(\frac{2L}{\tau_y}\right)^{1/4}e^{-\frac{y^2}{2l_B^2}}\theta {n/L \brack 0} (L z, L \tau) \,.
\ee
These are
eigenstates of the operator $\exp(\frac{2\pi i}{\tau_y} \pi_y)$, where
$\pi_y$  is the guiding center $y$-components.
For any $\tau$, the restriction of these orbitals to the real axis spans a different subspace
$\cL_\tau$ of the 1D Hilbert space $\cH$, which is in one-to-one correspondence
with the lowest Landau level at $\tau$.

To see why the orbitals $\chi_n$ are a natural choice of basis in the present context, 
we observe that the mapping
to the 1D Hilbert space $\cH$ introduces a new scalar product between
wave functions, defined as usual by integration over $[0,1]$ (instead of 
integration over the
fundamental domain in 2D).
\Eq{chin} as written is normalized independent of $n$
with respect to the 2D scalar product, but not with respect to the
scalar product of $\cH$.
However, these orbitals are orthogonal in both cases
thanks to trivial considerations of properties under
translation in $x$, which are unaffected by the 1D
mapping. 
The fact that the basis \Eq{chin} remains orthogonal,
and in particular linearly independent, after restriction to the real axis
makes it manifest that the mapping 
between the original lowest Landau level and its image $\cL_\tau$ 
in the 1D Hilbert space
is one-to-one.

We note that working with  $y$-guiding-center eigenstates
instead of $x$ (as in our initial discussion for the cylinder)
leaves the second quantized Hamiltonian invariant,
except for the trivial replacement 
$\tau \rightarrow -1/\tau$ associated with the ``modular S transformation'' .
This is  due to the ``S-duality''
of the physics on the torus (see, e.g., Ref. \onlinecite{seidel_duality}).
The torus Hamiltonian \eqref{V1_tor} was already written
with reference to the orbitals \eqref{chin}.

\subsection{Definition of a 2-body operator generating the deformation of guiding center variables}
\label{def_Gt}

We first explain how to relate a result obtained
within the 1D framework introduced above 
to the desired one, which uses ordinary conventions based
on a Hilbert space equipped with the standard 2D
scalar product.

\begin{figure}[tbh]
  \centering
    \includegraphics[width=0.45\textwidth]{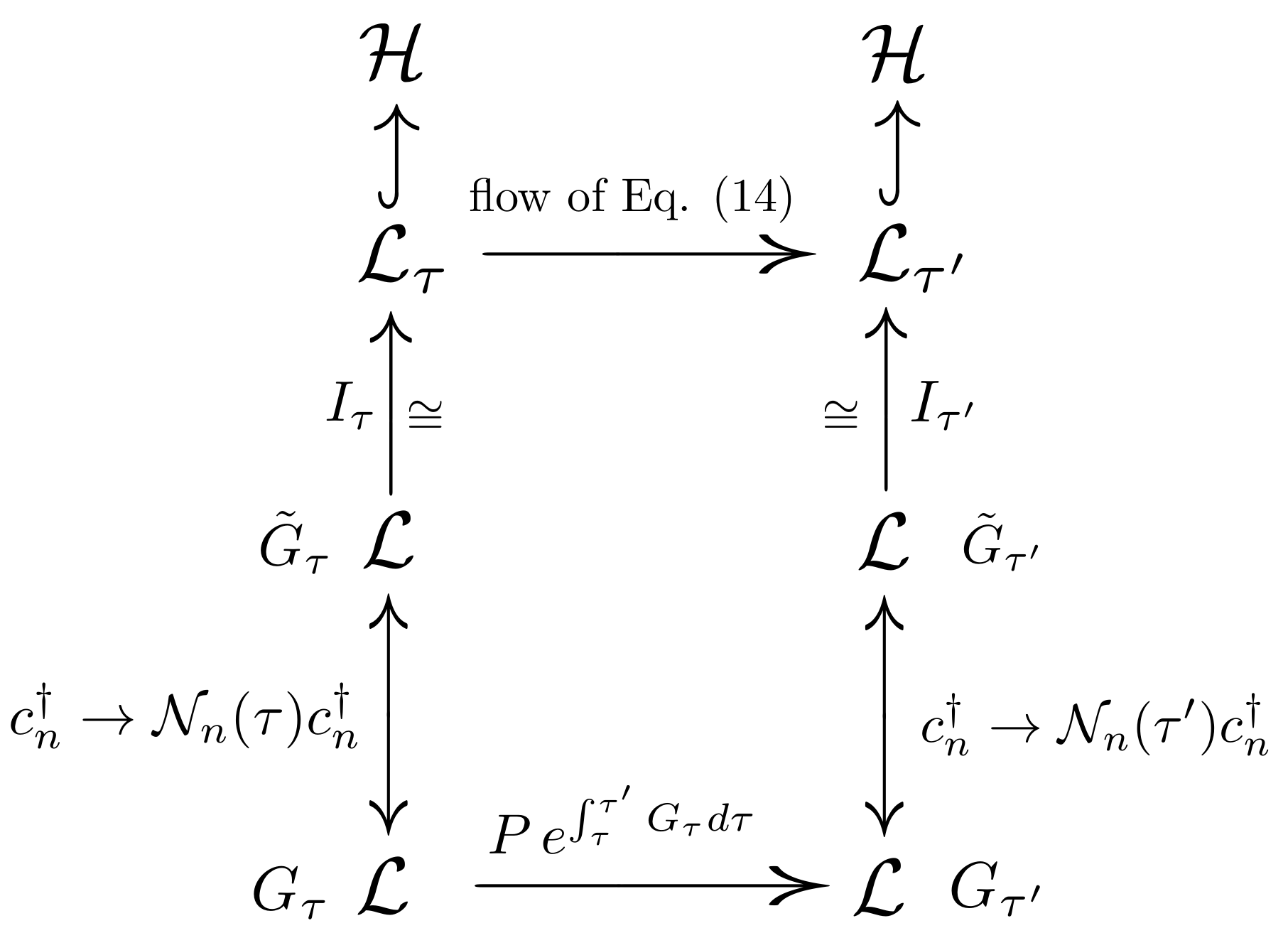}
     \caption{\label{commuting}
     Commuting diagram displaying the various Hilbert spaces and sub-spaces defined in the main text,
     and operators acting between them. The top segment shows lowest Landau levels $\cL_\tau$, $\cL_{\tau'}$ at different modular
     parameter that, using the 2D to 1D mapping defined in the text,  have been embedded into the same larger
     1D Hilbert space $\cH$. At the same time, each Landau level is isomorphic (through embeddings $I_\tau$, $I_{\tau'}$) 
     to the same finite dimensional
     ``abstract'' Landau level space $\cL$, in which only the guiding center degrees of freedom are represented. 
     The generator $\tilde G_{\tau}$ of changes in the guiding center degrees of freedom with $\tau$ is first constructed using the
     normalization conventions of the 1D Hilbert space. It is related by a similarity transformation to the operator
     $G_\tau$, which generates the analogous changes for the normalization convention of the usual 2D Hilbert space.
     In the horizontal direction, we have mappings between states defined for values of the modular parameter.
     The upper line is defined through the flow of \Eq{heat},
      which describes precisely the change of the Laughlin state,
     restricted to the real axis. The lower line represents the corresponding change in guiding center degrees of freedom,
     given by \Eq{integral}.
     The operator $G_\tau$ is constructed such that the diagram commutes. 
     }
\end{figure}

Suppose we have an operator $\tilde G_{\tau_x}$ ($\tilde G_{\tau_y}$)
that generates the change with $\tau_x$ ($\tau_y$)  in the coefficients $\tilde C_{\{n_k\}}(\tau)$
in the expansion of the Laughlin state,
\be\label{ltau}
 \psi_{1/q}(\tau) =\sum_{\{n_k\}}  \tilde C_{\{n_k\}}(\tau)\,  {\cal A} \, \tilde \chi_{n_1}(z_1,\tau)\cdot\dotsc\cdot   \tilde \chi_{n_N}(z_N,\tau),
\ee
where $ \tilde \chi_{n}(z,\tau)= {\cal N}_n(\tau)  \chi_{n}(z,\tau)$, ${\cal N}_n(\tau)$ being the factor that normalizes the state $\chi_n$
with respect to the 1D scalar product, $~_1\langle\phi|\psi\rangle _1= \int_0^1 dx\, \phi^\ast(x) \psi(x)$, i.e.,
${\cal N}_n=~_1\langle\chi_n|\chi_n\rangle _1^{-1/2}$, and we will often leave the $\tau$-dependence understood.
Likewise, we have dropped the label $\ell$ for now, which is just a spectator in the ``heat equation'' \eqref{heat}.
${\cal A}$ denotes anti-symmetrization in the indices $n_k$. The Laughlin state in
\Eq{ltau} is a member of the subspace $\cL_\tau$ of $\cH$ as defined in the preceding section.
We may now map the state \eqref{ltau} to the abstract Landau level Hilbert space $\cL$ as discussed in Sec. \ref{cylinder},
by applying a projector which ``forgets'' the degrees of freedom associated with kinetic momenta.
This situation is represented by the diagram in \Fig{commuting}.
If we perform this projection orthogonally with respect to the 1D scalar product, we obtain a ket
\be
    |\tilde \psi_{1/q}(\tau)\rangle =\sum_{\{n_k\}}  \tilde C_{\{n_k\}}(\tau)  c_{n_1}^\dagger\dotsc  c_{n_N}^\dagger |0\rangle\,.
\ee
By definition, we then have 
\be\label{Gtilde}
   \nabla_\tau |\tilde \psi_{1/q}(\tau)\rangle =  \tilde {\v G}_\tau |\tilde \psi_{1/q}(\tau)\rangle\,,
\ee
where we assume $\tilde {\v G}_\tau=(\tilde G_{\tau_x}, \tilde G_{\tau_y})$ to be of the form
\be\label{Gmmnn}
\tilde {\v G}_\tau= \sum_{mm'nn'} {\v G}_{mm'nn'}  c_m^\dagger  c_{m'}^\dagger  c_n c_{n'}\,.
\ee
In the end, one wants to do the projection of \Eq{ltau} orthogonally with respect to the original
2D scalar product. This gives
\be\label{Cnk}
    |\psi_{1/q}(\tau)\rangle =\sum_{\{n_k\}}   C_{\{n_k\}}(\tau)  c_{n_1}^\dagger\dotsc  c_{n_N}^\dagger |0\rangle
\ee
where $C_{\{n_k\}} = {\cal N}_{n_1}\dotsc  {\cal N}_{n_N}\tilde C_{\{n_k\}}$ from the change of normalization,
$c_n^\dagger\rightarrow {\cal N}_n c_n^\dagger$.
This implies the relation
\be\label{similarity}
    |\psi_{1/q}(\tau)\rangle=  e^{\sum_{n=0}^{L-1} \ln [{\cal N}_{n}(\tau)]  c^\dagger_n c_n } |\tilde \psi_{1/q}(\tau)\rangle\,.
\ee 
From this last line, we obtain that the desired operator
${\v G}_\tau$ defined by \Eq{dtau} 
is related to \Eq{Gmmnn} via \footnote{It turns out that the final form of $\tilde {\v G}_\tau$
also contains a one body part that we omit in \eqref{Gmmnn}, \eqref{Gtau1} for brevity.
However this part transforms analogously.}
\begin{align}
\label{Gtau1}
\nonumber {\v G}_\tau&= \sum_n \frac{\nabla_\tau \cN _n(\tau)}{\cN_n(\tau)} c^{\dagger}_n c_n\\
 & + \sum_{mm'nn'} \frac{\cN_m(\tau)\cN_{m'}(\tau)}{\cN_n(\tau)\cN_{n'}(\tau)} {\v G}_{mm'nn'} \, c_m^\dagger c_{m'}^\dagger c_n c_{n'}.
\end{align}
With this we have completely relegated the solution of the problem to the 1D Hilbert space.
We point out that the 1D mapping described above may generally provide an efficient way 
to calculate the matrix elements of operators acting within the lowest Landau level
on the torus.\footnote{We are indebted to G. M{\"o}ller for this observation.}
In this case, \Eq{Gtau1} will apply without the $\tau$-derivative part.
 The explicit form of $\cN_n (\tau)$ will be given below.
 
 We now define the operator $I_\tau$ which injects the ket $|\tilde \psi_{1/q}(\tau)\rangle$
 into ${\cL}_\tau\in \cH$, by sending $c^\dagger_n|0\rangle$ to $\tilde\chi_n(\tau)$.
 Thus
 \be\label{inject}
   \psi_{1/q}=I_\tau |\tilde\psi_{1/q}\rangle\,.
 \ee
 For the time being, we work at fixed $\tau_x$. Using
 the heat equation \eqref{heat} with $\partial_\tau=-i\partial_{\tau_y}$
 and differentiating \Eq{inject}, one obtains
\be\label{generator_y}
\partial_{\tau_y} \psi_{1/q}=i\Delta \psi_{1/q} = (\partial_{\tau_y} I_\tau) |\tilde \psi_{1/q}\rangle + I_\tau\,\tilde G_{\tau_y} |\tilde \psi_{1/q}\rangle \,,
\ee
where $\Delta$ denotes the differential operator on the right hand side of \Eq{heat}, and we also used \Eq{Gtilde}.


For $\mbox{Re}\,\tau=\mbox{Re}\,\tau'$,
 it is easy to see the $P_\tau \partial_{\tau_y} I_\tau\equiv 0$, where
$P_\tau$ is the orthogonal projection operator onto ${\cL}_\tau$
(we work in the 1D Hilbert space now, and will always refer to its scalar product when not stated otherwise).
To see this, it is sufficient to observe that $\langle  \chi_m | \partial_{\tau_y} \chi_n\rangle=0$ for all $m$, $n$.
This follows from the fact that $\langle  \chi_m(\tau) |  \chi_n(\tau')\rangle= \delta_{m,n}\langle  \chi_m(\tau) |  \chi_m(\tau')\rangle$
is always real for $\mbox{Re}\,\tau=\mbox{Re}\,\tau'$.
Thus, acting on the last equation with $P_\tau$, we get
\bd
P_\tau \Delta P_\tau \psi_{1/q} = P_\tau \Delta P_\tau I_\tau |\psi_{1/q}\rangle = I_\tau \tilde G_{\tau_y} |\tilde \psi_{1/q}\rangle \,.
\ed
where we have also inserted $P_\tau$ before $\psi_{1/q} \in {\cL}_\tau$.
Since we only
care about how the operator $\tilde G_{\tau_y}$ acts on
these $q$ states, for which we have the last equation,
we may thus define this operator
though the identity
\be
  \tilde G_{\tau_y} = I^{-1}_\tau  P_\tau \Delta P_\tau I_\tau\,.
\ee
The last equation expresses that the matrix elements of $\tilde G_{\tau_y}$
are just those of the differential operator $\Delta$ restricted to the
LLL subspace ${\cL}_\tau$. These can thus be calculated
straightforwardly by evaluating the standard expression for
two-body operators:
\be\label{Gmmnn_def}
G^y_{mm'nn'} =\frac{1}{2} \int_0^1 dx \int_0^1 dx' \,\tilde\chi_m^\ast(x)\tilde\chi_{m'}^\ast(x') \;\Delta\;  \tilde\chi_{n'}^\ast(x')\tilde\chi_{n}^\ast(x)\,.
\ee

As a last step, we calculate $G_{\tau_y}$ by fixing the normalization convention
for single particle orbitals in accordance
with the usual 2D scalar product, as displayed in \Eq{Gtau1}.
We may then obtain the generator for changes in $\tau_x$ simply
by studying the analytic properties of the coefficients $C_{\{n_k\}}(\tau)$ in \Eq{Cnk}.
As shown in Appendix \ref{appA}, one has
\be\label{Cxy}
 \partial_{\tau_x} C_{\{n_k\}}   = -i \partial_{\tau_y} C_{\{n_k\}}+i\frac{N}{4\tau_y}\,.
\ee
We can thus let $G_{\tau_x}=-i G_{\tau_y}+i\frac{\hat N}{4\tau_y}$.
Moreover, \Eq{Cxy} follows from the fact that $C_{\{n_k\}}/\tau_y^{N/4}$ is
holomorphic in $\tau$. We may use this insight to conveniently redefine
the normalization of the Laughlin states via
\be \label{newnorm}
   \psi_{1/q}'^\ell(z_1,\dotsc,z_N,\tau)= {\tau_y^{-N/4} } \,\psi_{1/q}^\ell(z_1,\dotsc,z_N,\tau)\,.
\ee 
The corresponding generator for changes in $\tau_y$ is then given by
$G'_{\tau_y}=G_{\tau_y}-\frac {\hat N}{4\tau_y}$. In the following,
we will always refer to the normalization convention \eqref{newnorm}.
Dropping all primes, we then have
\be\label{Grelations}
   G_{\tau_x}=-i G_{\tau_y}\equiv G_\tau\,.
\ee
With the ket $|\psi_{1/q}(\tau)\rangle$ now referring to \Eq{newnorm},
$|\psi_{1/q}(\tau)\rangle$ is then holomorphic in $\tau$, and we have
\be\label{Gholom}
   \partial_\tau | \psi_{1/q}(\tau)\rangle =  G_\tau |\psi_{1/q}(\tau)\rangle\,.
\ee

We present our final result as
\be
       G_\tau=G_0+ \frac{1}{4\pi iq}G_1+q G_2\;.
\ee
Here, the first term corresponds to $-i$ times the one-body operator in the $y$-component of \Eq{Gtau1},
plus the shift of $iN/4\tau_y$ shown in \Eq{Cxy}. 
Defining the functions
\be
  {\cal S}_n^a= \sum_l (2\pi i [lL+n])^ae^{-2\pi  L\tau_y (l+n/L)^2}\,,
\ee
the normalization factors defined above correspond to $(\cN_n)^{-2}= \sqrt{2L/\tau_y} \cS_n^0$.
We thus get
\be
  G_0=\frac{i}{2}\sum_n \frac{\partial_{\tau_y}{\cal S}^0_n}{{\cal S}_n^0}\,c_n^{\dagger}c_n=-\frac{1}{4\pi i L}\sum_n \frac{{\cal S}^2_n}{{\cal S}_n^0}\,c_n^{\dagger}c_n\;.
\ee
Next, $G_1$ is the contribution coming from the differential operator $\partial_X^2$ in \Eq{heat}.
Note that after normal ordering, the square of a single body operator still contains
a single body operator. We thus get the following result:
\be
  G_1=(\frac{q}{L})^2 [\sum_{n} \frac{{\cal S}^2_n}{{\cal S}^0_n} c_{n}^{\dagger}c_{n}+\sum_{n_1\neq n_2 }
\frac{{\cal S}^1_{n_1} {\cal S}^1_{n_2}}{{\cal S}^0_{n_1} {\cal S}^0_{n_2}} c_{n_1}^{\dagger}c_{n_1}c_{n_2}^{\dagger}c_{n_2}] \,.
\ee
Note that $ {\cal S}^2_n \neq  ({\cal S}^1_n)^2$, owing to the fact that
$P_\tau \partial_X^2 P_\tau \neq (P_\tau \partial_X P_\tau)^2$.
Finally, $G_2$ relates to the $\theta$-function part of \Eq{heat}.
\Eq{Gmmnn_def} can be evaluated by expanding the factors in the
integrand, which are all periodic in $x$, $x'$, into Fourier series.
For the $\partial\theta/\theta$-terms, this can be done by contour
integration and using known properties of $\theta$ functions.
Straightforward but tedious calculation allows one to 
express $G_2$ through rapidly converging, albeit multiple sums,
\be\label{G2}
G_2=  \frac{1}{2}\sum_{n_1n_2n_3n_4} \frac{\Delta_{n_1n_2n_3n_4}}{{\cal S}_{n_1}^0 {\cal S}_{n_2}^0} c_{n_1}^{\dagger}c_{n_2}^{\dagger}c_{n_4}c_{n_3}+\frac{1}{2} C\hat N(\hat N-1) \,,
\ee
and we have defined the function
\begin{equation}
   \begin{split}
          \Delta_{n_1n_2n_3n_4}&=\delta_{n_1+n_2,n_3+n_4} \frac{2\pi}{i}  \sum_{\substack{n\neq 0\\n=n_3-n_1 \mod L}} (\frac{e^{i\pi\tau n}}{1-e^{2i\pi\tau n}})^2 \\
&\sum_{l_1}e^{i\pi\tau L[(n_1+n)/L+l_1]^2}(e^{i\pi\tau L(n_1/L+l_1)^2})^*\\
 &\sum_{l_4}(e^{i\pi\tau L[(n_4+n)/L+l_4]^2})^* e^{i\pi\tau L(n_4/L+l_4)^2} \,,
   \end{split}
\end{equation}
and the ($\tau$-dependent) constant
\be\label{C}
  C=\frac{1}{4\pi i}[\int_{-1/2}^{1/2}(\frac{\partial_\tau\theta_4}{\theta_4})^2dz-\pi^2 ]\,.
\ee
In the above, the Kronecker $\delta$ is understood to be periodic, enforcing
identity $n_1+n_2=n_3+n_4 \mod\; L$.

\subsection{Symmetry considerations\label{symmetries}}

The operator $G_\tau$ defined in the preceding section 
is manifestly invariant under magnetic translations in the $x$-direction.
In the basis we chose here, this is tantamount to the conservation,
modulo $L$,
of the ``center-of-mass'' operator $\sum_n n c^\dagger_n c_n$. On the other hand, the operator is not
invariant under magnetic translations in the $\tau$-direction,
which amounts to an ordinary shift of the orbital indices. 
As already pointed out in Sec.
\ref{torus_intro}, the symmetrized operator
\be\label{Gtsymm}
G_{\tau,\mbox{\tiny sym}}=\frac{1}{L}\sum_{n=0}^{L-1} T_\tau^n \, G_\tau (T_\tau^\dagger)^n 
\ee
also satisfies \Eq{dtau}, where $T_\tau$ generates magnetic translations in $\tau$.
This is a trivial consequence of the fact that the $|\psi_{1/q}^\ell(\tau)\rangle$
transform among themselves under $T_\tau$, and all satisfy \Eq{dtau}.
Likewise, each term on the right hand side of \Eq{Gtsymm} satisfies \Eq{dtau}.
We may thus define the $L-1$ linearly independent 2-body operators
\be\label{Dn}
D_n= G_{\tau,\mbox{\tiny sym}}-T_\tau^n \,G_{\tau} (T_\tau^\dagger)^n,\quad n=0\dotsc L-2,
\ee
that all annihilate each of the $q$-fold degenerate Laughlin states,
\be
D_n |\psi_{1/q}^\ell\rangle = 0\;.
\ee
We note that the $D_n$ are not in any obvious way related to the operators $Q^\dagger_R Q_R$ of the pseudo-potential Hamiltonian, 
with $Q_R$ given by \eqref{V1_tor}. Indeed, the $D_n$ have a non-vanishing single-body term, whereas the $Q^\dagger_R Q_R$ do not.
The $D_n$ thus represent a new class of two-body operators that annihilate the torus Laughlin states (in the absence of quasi-holes).
For $q=3$ and various values of particle number $N$, we have verified that the property \eqref{Dn} characterizes the $q=3$ Laughlin states
uniquely.

Note that the single-body contribution to $G_{\tau,\mbox{\tiny sym}}$ is proportional to the particle number, as explained in Sec.
\ref{torus_intro}. 
This term can thus be replaced by a constant when acting on the Laughlin state, and hence can be ignored altogether
in practical calculations, where the real part of this constant is usually adjusted to fix the normalization of the
state (see below), and the imaginary part only affects the phase convention. For the same reason, we do not need
 the value of the $\tau$-dependent constant $C$ defined in \Eq{C} for the purpose of practical calculations.


\subsection{Presentation of the Laughlin state through its thin torus limit\label{present}}

In the following, we will generally identify $G_\tau$ with the symmetrized operator $G_{\tau,\mbox{\tiny sym}}$ discussed in the preceding section,
without carrying along the "sym" label. 
Putting the results of Sec. \ref{def_Gt} in integral form, we have, via \Eq{Gholom},
\be\label{integral}
    |\psi_{1/q}^\ell(\tau')\rangle = P\, e^{\int_{\tau}^{\tau'} G_\tau d\tau}  \, |\psi_{1/q}^\ell(\tau)\rangle \,,
\ee
where $P$ means path ordering. The integral in \Eq{integral} should be interpreted as a complex contour
integral, where the result is independent of the path connecting $\tau$ and $\tau'$.
This is so since by construction, $G_\tau$ generates exactly the change with $\tau$ of the guiding
center coordinates of the states in \Eq{newnorm}, which are single valued
functions of $\tau$. (This requires that we carry along all the $\tau$-dependent
c-number terms mentioned in the preceding section.)

We may also want to add, possibly different, real constants to $G_{\tau_x}$ and
$G_{\tau_y}$, such that the normalization of the Laughlin state is preserved
under the evolution with these operators. When evaluating \Eq{integral}
iteratively, this simply corresponds to normalizing the state at each step.
We denote the accordingly modified operators by $G_{\tau_x}^N$ and
$G_{\tau_y}^N$, and introduce the operator valued 1-form
$dG^N_\tau=G_{\tau_x}^N d\tau_x+G_{\tau_y}^N d\tau_y$.
We may then write
\be\label{integral_N}
    |\psi_{1/q}^\ell(\tau')\rangle_N = P\, e^{\int_{\tau}^{\tau'} dG_\tau ^N}  \, |\psi_{1/q}^\ell(\tau)\rangle_N \,,
\ee
where the subscript $N$ denotes normalized Laughlin states.
We are now interested in the thin torus limit $\tau\rightarrow i\infty$,
in which $ |\psi_{1/3}^\ell(\tau)\rangle_N$ approaches
the ket $|100100100\dotsc\rangle$,\cite{seidel1} or one related to the latter
through repeated action of $T_\tau$. Here, the labels 100100100\dots
are occupation numbers in the basis \eqref{chin}.
Given our earlier discussion for the cylinder,
it cannot be taken for granted that \Eq{integral_N} remains
well-defined in this limit. On the other hand, it may seem plausible that
this is the case, since the operators $G_{\tau_x}^N$,
$G_{\tau_y}^N$ do generate  off-diagonal matrix elements when acting on
the thin torus state, unlike the case of the cylinder. It thus seems feasible that
 the full Laughlin states at arbitrary $\tau$ admit the following presentation
 in terms of their respective thin torus limit,
 \be\label{presentation}
    |\psi_{1/q}^\ell(\tau)\rangle_N = P\, e^{\int_{\infty}^{\tau} dG_\tau ^N}  |...100...100...100...\dotsc\rangle\,,
 \ee
 where the pattern on the right hand side denotes one of the $q$ thin torus patterns at filling factor $1/q$.
The correctness of the above assertion remains non-trivial, however, as the $\tau'\rightarrow\infty$ limit
in \Eq{integral_N} must  be taken with care. In the next 
Section we provide numerical evidence for $q=3$, demonstrating the above
relation for various particle numbers $N$.
We thus find that the full torus Laughlin state may be generated 
 from its given thin torus limit
via application of the above path-ordered exponential involving the two-body operator constructed here.
We conjecture that this is true for general $q$. An application demonstrating this technique will be 
discussed in the following.

\section{Application: Hall viscosity
\label{viscosity}}

As an application of our findings in Sec. \ref{Gen}, we use \Eq{presentation}
(or the differential form \Eq{Gholom})
to calculate the $\nu=1/3$ torus Laughlin state along a contour in the complex $\tau$-plane, 
starting from the thin torus
limit at $\tau=i\infty$. As a physical motivation for calculating 
the Laughlin state along such contours, we will be asking how the
Hall viscosity\cite{read09} evolves along such contours. 
This quantity is naturally related to the main theme of 
of our paper, i.e., changes of the Laughlin state with changes in geometry.
The notion of a 
Hall viscosity of fractional quantum Hall liquids has  generated much interest recently,\cite{read09, haldane09, read_rezayi11}
expanding earlier work\cite{avron95} on integer quantum Hall states. 
In particular, in an insightful paper, \cite{read09} Read has
derived a general relation 
between the viscosity of a quantum Hall fluid and a characteristic
quantum number $\bar s$, which can be interpreted as ``orbital spin per particle'' and is related to the 
conformal field theory description of the state in question.
Here we only give a brief summary of the relevant definitions,
following closely Ref. \onlinecite{read_rezayi11}, to which we refer the interested reader for details.

We denote the fourth-rank viscosity tensor of the fluid by $\eta_{abcd}$, where we
are interested in the case of two spatial dimensions. 
In a situation with no dissipation, only its anti-symmetric or ``Hall viscosity'' component
$\eta^{(A)}_{abcd}=-\eta^{(A)}_{cdab}$ may be non-zero,
and this is possible only when time reversal symmetry
and the symmetry under reflection of space are both broken.
This is the situation in a magnetic field (where in a constant field,
only the product of these two symmetries is unbroken).

We now consider a system with periodic boundary
conditions defined by two periods $\v L_1=(L_1,0)$
and $\v L_2= (L_1\tau_x, L_1\tau_y)$, and Hamiltonian
\begin{equation}\label{Hg}
\begin{split}
H=\frac{1}{2}\sum_{i=1}^N& g^{ab} \pi_{ia}\pi_{ib}\,+\\
&P_{\mbox{\tiny LLL}} \,\sum_{m,n} \sum_{i<j} V(||x_{ij}+m{\v L}_1+n{\v L_2} ||_g)\, P_{\mbox{\tiny LLL}}
\end{split}
\end{equation}
Here, $\pi_a$ is a component of the kinetic momentum, $P_{\mbox{\tiny LLL}}$
denotes LLL-projection, and we have introduced a metric $g_{ab}$.
We have also introduced the ``periodized'' version of a potential
$V$ that depends on $x_i$, $x_j$ only via
$||x_{ij}||_g\equiv g_{ab}x_{ij}^ax_{ij}^b$ with $x_{ij}=x_i-x_j$.
We follow Ref. \onlinecite{read_rezayi11} and parametrize the
metric via $g(\lambda)=\Lambda^T \Lambda$, $\Lambda=\exp(\lambda)$, where $\Lambda$ can be
viewed as a coordinate transformation that transforms the identity metric into the
metric $g$. 
Clearly, $g$ is invariant under $\Lambda\rightarrow R\Lambda$, where $R$ is a 
rotation matrix. Since $\lambda$ can be interpreted as being proportional to
an ``infinitesimal version'' of $\Lambda$, whose rotational component is just its anti-symmetric
part, we may fix this rotational degree of freedom by requiring $\lambda$ to be symmetric.
Then, the Hall viscosity of
the ground state of \Eq{Hg}
 can be related\cite{avron95, read_rezayi11, read12} to
the adiabatic curvature on the space of background metrics,
here parameterized by the symmetric matrix $\lambda$.
Specializing to $g=\mbox{id}$, we have:
\be\label{eta}
\eta^{(A)}_{abcd}=-\frac{1}{V} F_{ab;cd}
\ee
where $F_{ab;cd}$ is the Berry curvature
\be
   F_{ab;cd}=   -2\mbox{Im}\langle \partial_{\lambda_{ab}} \psi | \partial_{\lambda_{cd}}\psi\rangle |_{g=\mbox{id}}\;,
\ee
and $\psi$ denotes the ground state of \Eq{Hg}.
$F_{ab;cd}$ clearly has the anti-symmetry of $\eta^{(A)}_{abcd}$, and it is also
symmetric in the index pairs $ab$ and $cd$.
Furthermore, at least in the thermodynamic limit of large $L_1$,
one would expect $\eta^{(A)}_{abcd}$ to acquire full rotational symmetry.
In two dimensions, this requires the trace $\eta^{(A)}_{abcc}$ to vanish,
where we use the sum convention, and similarly for the first index pair.
(In higher dimensions, rotational symmetry requires $\eta^{(A)}$ to vanish
identically). Moreover, in an incompressible fluid, the strain tensor $u_{ab}$
must be traceless. Therefore, since the viscosity couples to the rate of strain
$\dot u_{ab}$ via $\eta^{(A)}_{abcd}\dot u_{cd}$ to give a viscous contribution to the
stress tensor, only the traceless part of $\eta^{(A)}_{abcd}$ is of interest.
 It therefore makes sense to restrict our attention to traceless 
$\lambda_{ab}$, corresponding to volume preserving coordinate transformations.
Requiring  $F_{ab;cd}$ thus to be anti-symmetric, symmetric in the first and second pair, 
as well as traceless,
in $D=2$ the associated curvature 2-form  $F=\frac{1}{2}F_{ab;cd} \, d\lambda_{ab} \wedge
d\lambda_{cd}$ can only depend on the following two independent linear 
combinations of 1-forms, $d\lambda_{11}-d\lambda_{22}$ and 
$d\lambda_{12}+d\lambda_{21}$. Hence it must be proportional
to their product:\cite{read_rezayi11}
\be\label{F}
   F= -\frac{1}{2} s\, (d\lambda_{11}-d\lambda_{22})\wedge (d\lambda_{12}+d\lambda_{21})\,,
\ee
and we introduced a proportionality factor $-s/2$ whose physical meaning
will be given below. The above expression in \Eq{eta}
gives
\be \label{eta2}
\eta^{(A)}_{abcd}=\eta^{(A)}( \delta_{ad}\epsilon_{bc}+\delta_{bc}\epsilon_{ad})
\ee
with
\be
 \eta^{(A)}=\frac{1}{2} \bar s\bar n\hbar\;,
\ee
where $\bar n=N/V$ is the particle density, $\bar s=s/N$, and we have
restored a factor of $\hbar$. 
As shown in Ref. \onlinecite{read09},
 in the thermodynamic limit the parameter $\bar s$ 
is quantized and
can be identified with the average orbital spin per particle, which is
related to the conformal dimension of the field describing
particles in the conformal field theory description of the state.
It is further related to the topological shift on the sphere, $\cal S$,  of the underlying
state via $\bar s={\cal S}/2$. For the Laughlin $1/3$ state,
$\bar s =3/2$.

We now consider fixed boundary conditions described by $\tau$, and introduce a metric
that corresponds to the infinitesimal transformation
\be
d\lambda= \frac{1}{2\tau_y} \begin{pmatrix}
- d\tau_y & d\tau_x \\
d\tau_x  & d\tau_y
\end{pmatrix}\,.
\ee
It is not difficult to see that the corresponding metric change
is equivalent to changing the modular parameter $\tau$ to
$\tau'=\tau+d\tau_x+i d\tau_y$.
We may thus rewrite \Eq{F} as
\be\label{Ftau}
   F=-\frac{N\bar s}{2\tau_y^2}\, d\tau_x\wedge d\tau_y\,.
\ee
To each $\lambda$ can be associated a $\tau'$, where
$\Lambda=\exp(\lambda)$ is the coordinate transformation
that changes the $\tau$-boundary condition into a $\tau'$-boundary
condition, where
\be\label{taup}
   \tau'= \frac{\Lambda \v L_1\cdot\Lambda \v L_2 + i\Lambda \v L_1 \times \Lambda \v L_2 }{||\Lambda \v L_1||^2}\,.
\ee
For fixed $\tau$, we now parameterize $\lambda$, and thus the metric, by $\tau'$.
(Note that the right hand side of \Eq{taup} can be viewed as a function of $\Lambda$ and $\tau$.)
\Eq{Ftau} then implies that
\be\label{sbar}
     \bar s = \frac{4\tau_y^2}{N}\, \mbox{Im}\, \langle \partial_{\tau'_x} \psi | \partial_{\tau'_y}\psi \rangle\,.
\ee
We emphasize that in the above, $\psi\equiv \psi(\tau,g_\tau(\tau'))$ always satisfies the same
boundary condition defined by $\tau$, and depends on $\tau'$ only through the metric.
At the same time, $\psi(\tau,g_\tau(\tau'))$ is related to $\psi(\tau',\mbox{id})$
by the unitary transformation $\chi_n(\tau,g_\tau(\tau'))\rightarrow\chi_n(\tau',\mbox{id})$,
with $\chi_{n}(\tau, g(\tau'))$ the deformed version of the state \eqref{chin} in the presence
of the metric $g(\tau')$.
However, for fixed $\tau$, the $\psi(\tau, g_\tau(\tau'))$  live in the same Hilbert space,\cite{read_rezayi11}
independent of $\tau'$.
The advantage of introducing both $\tau$ and $\tau'$, where the former describes boundary
conditions, and the latter describes the ``true geometry'' of the system, is that
 we may restrict ourselves to metrics $g_\tau(\tau')$
in the vicinity of the identity (corresponding to $\tau'$ close to $\tau$), such that
\Eq{eta} is directly applicable.

We now consider $\psi=\psi^N_{1/3}$, the 
normalized Laughlin $1/3$ state (where we suppress labels $\tau$, $\tau'$, and $\ell$).
We have the expansion 
\be\label{psiN}
\psi_{1/3}^N= \sum_{\{n_k\}} C^N_{\{n_k\}}(\tau') |{\{n_k\}}\rangle_g\,,
\ee
and $|{\{n_k\}}\rangle_g\rangle$ is short for the Slater determinant
${\cal A}\,\left[ \chi_{n_1}(z_1,\tau, g(\tau'))\cdot\dotsc\cdot   \chi_{n_N}(z_N,\tau, g(\tau'))\right]$.
 We write
$
-2\mbox{Im}\, \langle \partial_{\tau'_x} \psi | \partial_{\tau'_y}\psi \rangle=\nabla_{\tau'}\times A
$
where
\be\label{A}
A=i\sum_{\{n_k\}} \, \left(|C^N_{\{n_k\}}|^2~_g\langle{\{n_k\}}|\nabla_{\tau'}|{\{n_k\}}\rangle_g+ {C^N}^*_{\!\!\!\!\!\!\{n_k\}}\nabla_{\tau'} C^N_{\{n_k\}}\right)
\ee
is the Berry connection.
It turns out that in the first term, which describes the change of the LLL basis with $\tau'$,
$~_g\langle{\{n_k\}}|\nabla_{\tau'}|{\{n_k\}}\rangle_g$ is independent of $\{n_k\}$,
and contributes a constant $1/2$ to \Eq{sbar}.\cite{levay95}
The second term depends on the changes of the $C_{\{n_k\}}$ with $\tau'$, which we 
described in the preceding section. We first assume the general situation
where this change is described by \Eq{dtau} with two generators
$G_{\tau_x}$ and $G_{\tau_y}$ that are not necessarily related
and that do not necessarily preserve the normalization of the state.
It is straightforward to show that the contribution from the second term
then leads to the following connected expectation value,
\begin{equation}
\label{imcurl}
\begin{split}
  & -2\,\mbox{Im} \sum_{\{n_k\}}  \partial_{\tau'_x}{C^N}^\ast_{\!\!\!\!\!\!\{n_k\}}\partial_{\tau'_y} C^N_{\{n_k\}}\\
  & = i\left[\langle G_{\tau_x}^\dagger G_{\tau_y} - G_{\tau_y}^\dagger G_{\tau_x}  \rangle
      - \langle G_{\tau_x}^\dagger \rangle\langle G_{\tau_y}\rangle + \langle G_{\tau_y}^\dagger \rangle\langle G_{\tau_x}  \rangle
\right]\,,
\end{split}
\end{equation}
\begin{figure}[t!]
  \centering
    \includegraphics[width=0.48\textwidth]{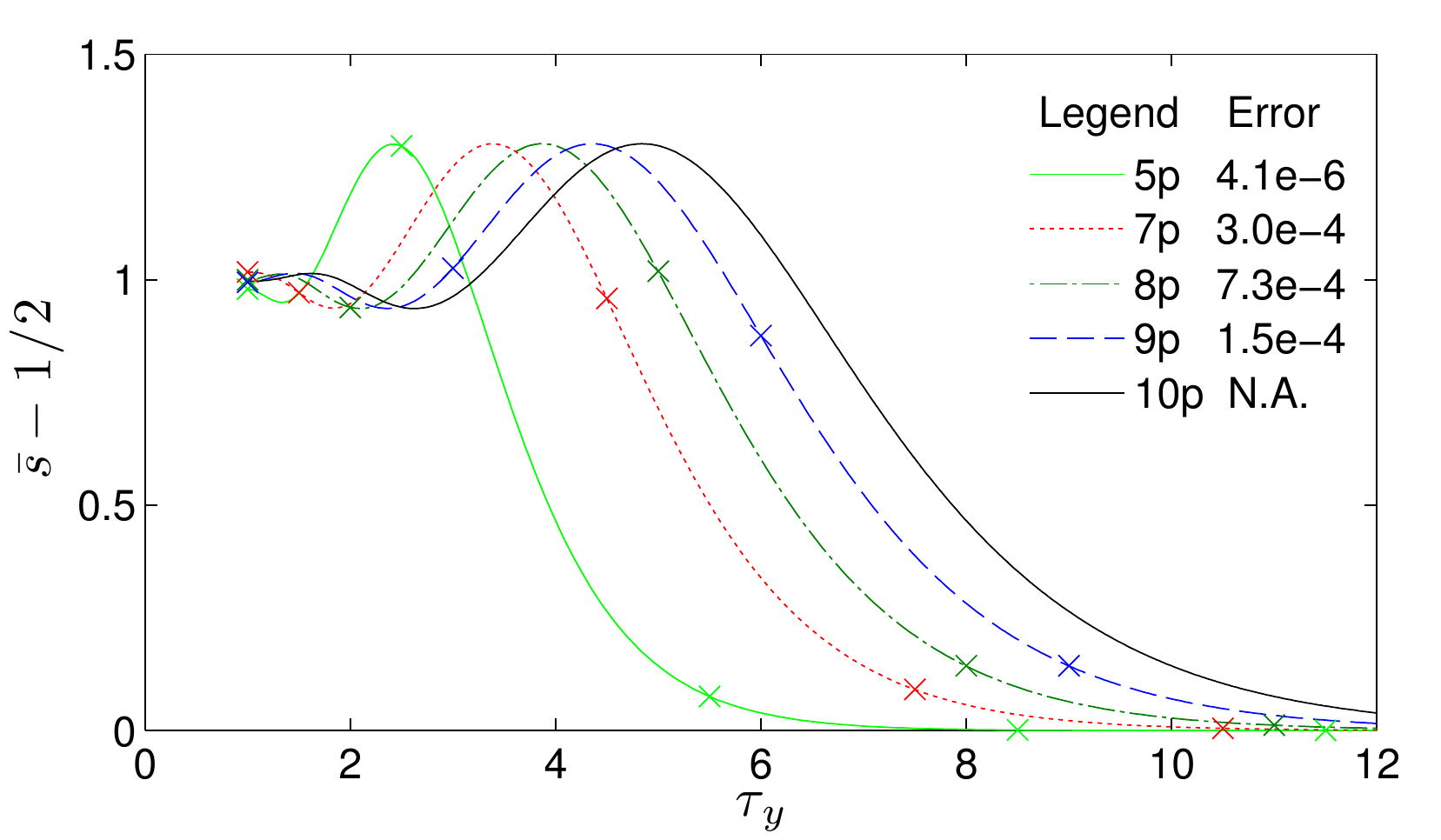}
     \caption{\label{viscosity1}Average ``orbital spin per particle'' $\bar{s}$ as calculated from \Eq{sbar},
      for the $\nu=1/3$ torus Laughlin state at $\tau$ generated via \Eq{presentation2}, using the procedure described in the main text and below.
     We start with the thin torus state at relatively large but finite $\tau'$, and iteratively solve \Eq{Gholom} using the 4th order Runge-Kutta method.
     Final state errors compared with exact diagonalization at $\tau=i$ are shown for 5 particles for $\tau'=30i$ and step size $d\tau=0.01i$, 7 and 8 particles for $\tau'=30i$ 
     and $d\tau=0.05i$, 9 particles for  $\tau'=40i$, $d\tau=0.025i$. 10 particles data is shown for $\tau'=80i$, $d\tau=0.02i$. 
   Crosses denote the value of $\bar{s}-1/2$  obtained from the exactly diagonalized Laughlin state in \Eq{sbar}, for comparison. The errors of $\bar{s}$ are at or smaller than $10^{-6}$ in these cases. All the errors decrease further with larger initial $\tau'$ and smaller step size.}
\end{figure}
where expectation values on the right hand side are taken in the state \Eq{psiN}.
The last two terms take care of the normalization, and will cancel if both
operators are anti-Hermitian (describing unitary evolution), in which case the
expression reduces to the expectation value of a commutator. Note also that
the expression is invariant under constant shifts of any of the two operators.
We now specialize to the case where these operators are related by \Eq{Grelations}.
 Plugging \Eq{A}, \Eq{imcurl} into \Eq{sbar}, this gives
\be\label{sbar2}
  \bar s= \frac{1}{2}+ \frac{4\tau_y^2}{N}|\Delta G_\tau|^2\,,
\ee
where $|\Delta G_\tau|= (\langle G_\tau^\dagger G_\tau\rangle -\langle G_\tau^\dagger \rangle \langle G_\tau \rangle)^{1/2}$
is the variance of the operator $G_\tau$ in the state $\psi^N_{1/3}$, and is manifestly positive
(the Laughlin state at $\tau$ certainly being
no eigenstate of $G_\tau$ for any $\tau$). As stated above, for the Laughlin
state $\bar s$ is expected to approach 
$3/2$ in the thermodynamic limit. This has been checked in Ref. \onlinecite{read_rezayi11},
by calculating torus Laughlin (and other) states by exact diagonalization of parent Hamiltonians,
and computing the Berry curvature by taking overlaps between such states for different
$\tau$ (or $\lambda$). 
Here we will consider the same problem both as a demonstration
and a consistency check of the results presented in the preceding section.
To this end, we calculate the Laughlin state from the presentation \eqref{presentation},
or by numerically integrating the differential equation \eqref{Gholom} with thin torus
initial conditions, and then computing $\bar s$ from \Eq{sbar2}.
Note that both steps of the calculation make use of the two-body operator $G_\tau$.
In particular, our results will confirm the accuracy of \Eq{presentation}, which may be written more carefully as
 \be\label{presentation2}
 |\psi_{1/q}^\ell(\tau)\rangle_N =\lim_{\tau'\rightarrow i\infty} P\, e^{\int_{\tau'}^{\tau} dG_\tau ^N}  |...100...100...100...\dotsc\rangle\,.
 \ee
Evaluating the expression on the right for some large but finite $\tau'$ 
is equivalent to integrating \Eq{Gholom} (and normalizing the result), 
where the thin torus limiting
state defines the initial condition at $\tau'$. This obviously
introduces some error compared to the full Laughlin state
at the initial value $\tau'$, hence also at the final value $\tau$.
Since it is not clear a priori how this error behaves in the limit
of large imaginary $\tau'$, possible pitfalls are that the limit
in \Eq{presentation2} is ill-defined, or that it is well-defined
but does not agree with the Laughlin state. \footnote{The latter
case is obviously realized for the Laughlin state on the cylinder
and the operator $G_{r^{-2}}$ defined in Sec. \ref{cylinder},
where the expression analogous to the right hand side \Eq{presentation2} leaves the
thin torus limiting state invariant.} 
Our results, however, give strong support of \Eq{presentation2}.

\begin{figure}
  \centering
    \includegraphics[width=0.48\textwidth]{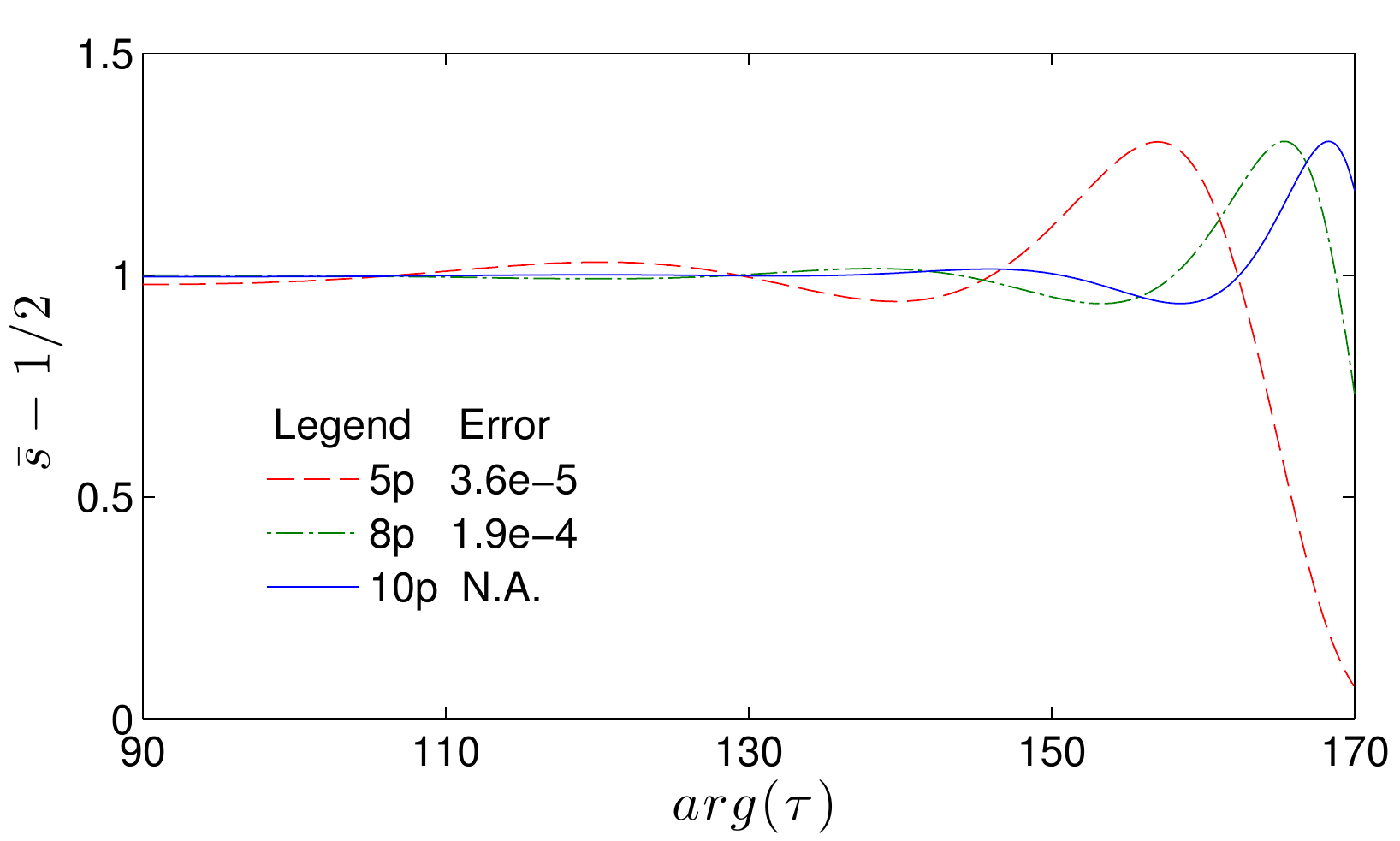}
     \caption{\label{viscosity2}Average orbital spin per particle $\bar{s}$,
      calculated from \Eq{sbar}, for the $\nu=1/3$
      torus Laughlin state. The state has been evolved out of the thin torus limit
      first down to $\tau=i$ as described in the caption of \Fig{viscosity1},
      and then to $\tau=e^{i\theta}$ using the same method.
  The step sizes used are  $d\theta=0.01rad$ for 5 particles and $d\theta=0.001rad$ for 8 and 10 particles. 
  The state difference with the exactly diagonalized Laughlin state at the last step is listed in the figure.
  Data stay close to the value $\bar s -1/2=1$ expected in the thermodynamic limit\cite{read09}
   for a wide range of angles $\theta$, as first observed in Ref. \onlinecite{read_rezayi11} from
   exact diagonalization.}
\end{figure}

Fig. \ref{viscosity1} shows the results for the value of $\bar s-1/2$ from this method for $q=3$. Beginning with the thin torus state $|100100100\dotsc\rangle$ at large imaginary $\tau'$,
we evolve the state down to $\tau=i$, i.e., a torus of aspect ratio 1,
integrating \Eq{Gholom} using the classical 4th order Runge-Kutta method.
We normalize the state at each step.
For particle numbers $N=5$ to $N=9$, we have observed that
the error of the state obtained at $\tau=i$, compared to
the Laughlin state generated from exact diagonalization of the 
$V_1$ Haldane pseudopotential,  $| \psi-\psi_{ed} |$, 
 becomes systematically
smaller with increasing initial $\tau'$ and decreasing step size $d\tau$.
The observed state error at $\tau=i$ has been on the order of $10^{-6}$
for $N=5$, $d\tau=0.01i$, and $\tau'=30$, and on the order of 
$10^{-4}$
for $N=9$, $d\tau=0.025i$, and $\tau'=40$. For $N=10$, we show data based on our method only. Generally, larger $N$ requires
larger $\tau'$ for the same accuracy. 
\Fig{viscosity1} shows $\bar s -1/2$  for various $N$ using our method, whereas crosses denote isolated points for which values have been obtained from exact diagonalization for comparison.
One sees that the expected value of $\bar s-1/2=1$ is always approached rather closely for $\tau=i$,
though it deviates from this value for $|\tau|$ noticeably larger than $1$. The crossover where notable deviations
from $1$
set in is pushed to larger $|\tau|$ with increased particle number, as expected.
However, the value of $\bar s$ is found to be much more constant, and close to 
its expected thermodynamic limit, when instead of varying the modulus of 
$\tau$ we vary its phase at $|\tau|=1$, even for five particles, 
as shown in \Fig{viscosity2}.
Data were obtained by continued integration of \Eq{Gholom} away
from $\tau=i$ along a contour where $\tau=\exp(i\theta)$.
$\bar s -1/2$ remains close to $1$ except for angles $\theta$
approaching $\pi$. These observations are consistent
with the exact diagonalization data published in Ref. \onlinecite{read_rezayi11}
for $N=10$ and (at $\tau=\exp(i\pi/3)$) larger particle number.

\section{Discussion\label{discussion}}

In the preceding sections, we considered the change 
in the guiding
center variables,
with modular parameter
$\tau$ for the torus Laughlin states.
Within a given Landau level, 
the guiding center coordinates fully specify the state.
We have shown that this change is generated by a two-body
operator $G_\tau$, which we have explicitly constructed.
We have demonstrated numerically that
by means of this two-body operator, the Laughlin state
for any modular parameter $\tau$ can be generated
from its simple thin torus ($\tau=i\infty$) limit.
The ability to generate the full torus
Laughlin state in this way may be compared 
to squeezing rules that follow from the
Jack polynomial structure of this state 
in other geometries.\cite{bernevig08, bernevig08-2}
 From a practical point of view, however,
our method still requires integration of
a first order differential equation.
While this requires some compromise between
accuracy and computational effort,
the added benefit is that in the process of
the calculation, the Laughlin state is generated
along an entire contour in the complex $\tau$ plane,
rather than just for a single value of $\tau$.
It is thus likely that whenever a moderate error
can be tolerated, but many $\tau$ values are of interest,
our method may become competitive
compared to numerical diagonalization.
As a demonstration of these features, we have produced
results relating to the Hall viscosity that are similar to
those of Ref. \onlinecite{read_rezayi11} (and are expected
to be identical within numerical accuracy for identical
particle number, which we have not yet studied).
The Hall viscosity is itself deeply related to our main
theme of study, i.e., geometric changes in the Laughlin state,\cite{avron95,read09,read_rezayi11}
and we have discussed its precise relation to the
generator constructed here (\Eq{sbar2}), following Ref. \onlinecite{read_rezayi11}.

We note that one key ingredient of our procedure is to 
embed different torus Laughlin states, which are related to one another
by the application of strain, into the same Hilbert space.
For this we make use of a dimensional reduction that is
made possible by the analytic properties 
of lowest Landau level wave functions on the torus.
We argued that this mapping may be useful in other contexts.
However, recent work on Hall viscosity\cite{read_rezayi11} achieves
the same embedding by a different method, which is to introduce
a metric describing the effect of strain, rather than a change in boundary
conditions. We conjecture that if we had used this method in Sec. \ref{Gen}, we would
have directly obtained the symmetrized version of our operator $G_\tau$.
In this way, however, we would not have obtained the family
of two-body operators given in Sec. \ref{symmetries}, whose members annihilate 
the torus Laughlin states.

While primarily, we have been working in a finite dimensional
Hilbert space that represents guiding center-coordinates only,
the operator defined in Sec. \ref{Gen} also naturally acts within the full Hilbert space,
which can be viewed as the tensor product of the degrees of freedom
for the guiding centers and the dynamical momenta, respectively.
Within this larger, physical, Hilbert space, the operator $G_\tau$
generates the change in guiding center degrees of freedom
associated to a change in the torus geometry, but not the corresponding
change of the Landau level. As pointed out recently by
Haldane,\cite{haldane11} the Laughlin state may be generalized
by the introduction of a geometric parameter that describes 
the deformation of guiding center variables in response to a
change in the ``interaction metric''. The so deformed Laughlin state
is still the exact ground state of an appropriately deformed Hamiltonian.
The operator that we have constructed can thus also be viewed as
generating the change of the torus Laughlin state in response
to a change of the interaction metric, i.e., the change in ground state
for the corresponding family of deformed pseudo-potential Hamiltonians.
For the disc geometry, this problem has been addressed from different angles
previously.\cite{read_rezayi11,kunyang12}

We conjecture that the observations made here are not limited to Laughlin states,
but can be generalized to other quantum Hall states as well. Indeed, a great wealth
of model wave functions is obtained from conformal blocks in rational conformal
field theories.\cite{MR} For conformal blocks on the torus, the dependence on the modular 
parameter $\tau$ can be described by Knizhnik-Zamolodchikov-Bernard (KZB) type
equations.\cite{bernard88} We expect therefore that our approach can be generalized to
other trial wave functions related to conformal field theories. The details of such generalizations
are left for future work.

\section{Conclusion\label{conclusion}}

In this work, we have shown that geometric changes in the guiding center coordinates
of the torus Laughlin state are generated by a two-body operator.
We have demonstrated that the equation that governs the evolution of the torus Laughlin state
as a function of the modular parameter $\tau$ can be continued into the thin torus
limit. This gives rise to a new presentation of the torus Laughlin state in its second quantized, or guiding
center, form. This presentation allows one to calculate the torus Laughlin states in terms of
a simple thin torus or ``dominance'' pattern by means of integration of the flow generated by the two-body operator
defined in this work. 
This operator hence realizes the adiabatic evolution of the simple thin torus product state
into the full Laughlin state on regular tori.
To demonstrate this, we have numerically compared
both the Laughlin state generated from this method,
as well as the Hall viscosity derived from it,  to exact diagonalization results.
While the demonstration of our new presentation of the torus Laughlin state rests in part on numerics,
we defer more detailed analytic studies to future investigation.

\begin{acknowledgments}
This work has been supported by the 
 National Science Foundation under NSF Grant No. DMR-1206781 ( ZZ and AS),
 and NSF Grant No. DMR-1106293 (ZN).
 AS would like to thank N. Read, K. Yang, I. Gruzberg, T.H. Hansson, and G. M\"oller
 for insightful comments. 
\end{acknowledgments}

\appendix

\section{Analytic properties of coefficients\label{appA}}

For definiteness, we will refer to the Laughlin state $\psi_{1/q}^\ell$
using the normalization conventions
 \eqref{laugh_tor}, \eqref{CM}.
The coefficients $C_{\{n_k\}}(\tau)$ defined in \Eq{Cnk}
then imply the following expansion of the analytic Laughlin state,
\be\label{expansion}
 \psi^\ell _{1/q}(\tau)=\sum_{\{n_k\}}   C_{\{n_k\}}(\tau)\,  {\cal A} \,  \chi_{n_1}(z_1,\tau)\cdot\dotsc\cdot    \chi_{n_N}(z_N,\tau)\,.
\ee
Here, as before, the symbol $\cal A$ denotes anti-symmetrization, and single particle orbitals $\chi_n$ 
are defined in \Eq{chin}. We define new orbitals $\chi_n'(\tau)=\tau_y^{1/4}\chi_n(\tau)$ that are holomorphic in $\tau$,
as is the Laughlin state $\psi_{1/q}^\ell(\tau)$. Hence, by acting with $\partial_{\bar \tau}=\frac{1}{2}(\partial_{\tau_x}+i\partial_{\tau_y})$
on \Eq{expansion}, we obtain
\be
0 =\sum_{\{n_k\}}  \left[\partial_{\bar \tau} \left(C_{\{n_k\}}(\tau)/\tau_y^{N/4}\right)\right]\,  {\cal A} \,  \chi'_{n_1}(z_1,\tau)\cdot\dotsc\cdot    \chi'_{n_N}(z_N,\tau)\,.
\ee
The linear independence of the orbitals $ \chi_n'(\tau)$ and of the associated many-particle Slater determinants then implies
\be\label{Cproperty}
\partial_{\bar \tau} \left(C_{\{n_k\}}(\tau)/\tau_y^{N/4}\right)=0\,,
\ee
i.e.,  the quantities $C_{\{n_k\}}(\tau)/\tau_y^{N/4}$ are holomorphic in $\tau$.
\Eq{Cxy} follows immediately from \Eq{Cproperty}.

\bibliography{bibheat-1}

\begin{thebibliography}{47}%
\makeatletter
\providecommand \@ifxundefined [1]{%
 \@ifx{#1\undefined}
}%
\providecommand \@ifnum [1]{%
 \ifnum #1\expandafter \@firstoftwo
 \else \expandafter \@secondoftwo
 \fi
}%
\providecommand \@ifx [1]{%
 \ifx #1\expandafter \@firstoftwo
 \else \expandafter \@secondoftwo
 \fi
}%
\providecommand \natexlab [1]{#1}%
\providecommand \enquote  [1]{``#1''}%
\providecommand \bibnamefont  [1]{#1}%
\providecommand \bibfnamefont [1]{#1}%
\providecommand \citenamefont [1]{#1}%
\providecommand \href@noop [0]{\@secondoftwo}%
\providecommand \href [0]{\begingroup \@sanitize@url \@href}%
\providecommand \@href[1]{\@@startlink{#1}\@@href}%
\providecommand \@@href[1]{\endgroup#1\@@endlink}%
\providecommand \@sanitize@url [0]{\catcode `\\12\catcode `\$12\catcode
  `\&12\catcode `\#12\catcode `\^12\catcode `\_12\catcode `\%12\relax}%
\providecommand \@@startlink[1]{}%
\providecommand \@@endlink[0]{}%
\providecommand \url  [0]{\begingroup\@sanitize@url \@url }%
\providecommand \@url [1]{\endgroup\@href {#1}{\urlprefix }}%
\providecommand \urlprefix  [0]{URL }%
\providecommand \Eprint [0]{\href }%
\providecommand \doibase [0]{http://dx.doi.org/}%
\providecommand \selectlanguage [0]{\@gobble}%
\providecommand \bibinfo  [0]{\@secondoftwo}%
\providecommand \bibfield  [0]{\@secondoftwo}%
\providecommand \translation [1]{[#1]}%
\providecommand \BibitemOpen [0]{}%
\providecommand \bibitemStop [0]{}%
\providecommand \bibitemNoStop [0]{.\EOS\space}%
\providecommand \EOS [0]{\spacefactor3000\relax}%
\providecommand \BibitemShut  [1]{\csname bibitem#1\endcsname}%
\let\auto@bib@innerbib\@empty
\bibitem [{\citenamefont {Tsui}\ \emph {et~al.}(1982)\citenamefont {Tsui},
  \citenamefont {Stormer},\ and\ \citenamefont {Gossard}}]{stormer82}%
  \BibitemOpen
  \bibfield  {author} {\bibinfo {author} {\bibfnamefont {D.~C.}\ \bibnamefont
  {Tsui}}, \bibinfo {author} {\bibfnamefont {H.~L.}\ \bibnamefont {Stormer}}, \
  and\ \bibinfo {author} {\bibfnamefont {A.~C.}\ \bibnamefont {Gossard}},\
  }\href {\doibase 10.1103/PhysRevLett.48.1559} {\bibfield  {journal} {\bibinfo
   {journal} {Physical Review Letters}\ }\textbf {\bibinfo {volume} {48}},\
  \bibinfo {pages} {1559} (\bibinfo {year} {1982})}\BibitemShut {NoStop}%
\bibitem [{\citenamefont {Laughlin}(1983)}]{laughlin}%
  \BibitemOpen
  \bibfield  {author} {\bibinfo {author} {\bibfnamefont {R.~B.}\ \bibnamefont
  {Laughlin}},\ }\href {\doibase 10.1103/PhysRevLett.50.1395} {\bibfield
  {journal} {\bibinfo  {journal} {Physical Review Letters}\ }\textbf {\bibinfo
  {volume} {50}},\ \bibinfo {pages} {1395} (\bibinfo {year}
  {1983})}\BibitemShut {NoStop}%
\bibitem [{\citenamefont {Moore}\ and\ \citenamefont {Read}(1991)}]{MR}%
  \BibitemOpen
  \bibfield  {author} {\bibinfo {author} {\bibfnamefont {G.}~\bibnamefont
  {Moore}}\ and\ \bibinfo {author} {\bibfnamefont {N.}~\bibnamefont {Read}},\
  }\href {\doibase 10.1016/0550-3213(91)90407-O} {\bibfield  {journal}
  {\bibinfo  {journal} {Nuclear Physics B}\ }\textbf {\bibinfo {volume}
  {360}},\ \bibinfo {pages} {362} (\bibinfo {year} {1991})}\BibitemShut
  {NoStop}%
\bibitem [{\citenamefont {Read}\ and\ \citenamefont {Rezayi}(1999)}]{RR}%
  \BibitemOpen
  \bibfield  {author} {\bibinfo {author} {\bibfnamefont {N.}~\bibnamefont
  {Read}}\ and\ \bibinfo {author} {\bibfnamefont {E.}~\bibnamefont {Rezayi}},\
  }\href {\doibase 10.1103/PhysRevB.59.8084} {\bibfield  {journal} {\bibinfo
  {journal} {Physical Review B}\ }\textbf {\bibinfo {volume} {59}},\ \bibinfo
  {pages} {8084} (\bibinfo {year} {1999})}\BibitemShut {NoStop}%
\bibitem [{\citenamefont {Haldane}\ and\ \citenamefont
  {Rezayi}(1988)}]{haldane_rezayi88}%
  \BibitemOpen
  \bibfield  {author} {\bibinfo {author} {\bibfnamefont {F.~D.~M.}\
  \bibnamefont {Haldane}}\ and\ \bibinfo {author} {\bibfnamefont {E.~H.}\
  \bibnamefont {Rezayi}},\ }\href {\doibase 10.1103/PhysRevLett.60.956}
  {\bibfield  {journal} {\bibinfo  {journal} {Physical Review Letters}\
  }\textbf {\bibinfo {volume} {60}},\ \bibinfo {pages} {956} (\bibinfo {year}
  {1988})}\BibitemShut {NoStop}%
\bibitem [{\citenamefont {Halperin}(1983)}]{halperin83}%
  \BibitemOpen
  \bibfield  {author} {\bibinfo {author} {\bibfnamefont {B.}~\bibnamefont
  {Halperin}},\ }\href {\doibase 10.5169/seals-115362} {\bibfield  {journal}
  {\bibinfo  {journal} {Helvetica Physica Acta}\ }\textbf {\bibinfo {volume}
  {56}},\ \bibinfo {pages} {75} (\bibinfo {year} {1983})}\BibitemShut {NoStop}%
\bibitem [{\citenamefont {Simon}\ \emph {et~al.}(2007)\citenamefont {Simon},
  \citenamefont {Rezayi}, \citenamefont {Cooper},\ and\ \citenamefont
  {Berdnikov}}]{gaffnian}%
  \BibitemOpen
  \bibfield  {author} {\bibinfo {author} {\bibfnamefont {S.~H.}\ \bibnamefont
  {Simon}}, \bibinfo {author} {\bibfnamefont {E.~H.}\ \bibnamefont {Rezayi}},
  \bibinfo {author} {\bibfnamefont {N.~R.}\ \bibnamefont {Cooper}}, \ and\
  \bibinfo {author} {\bibfnamefont {I.}~\bibnamefont {Berdnikov}},\ }\href
  {\doibase 10.1103/PhysRevB.75.075317} {\bibfield  {journal} {\bibinfo
  {journal} {Physical Review B}\ }\textbf {\bibinfo {volume} {75}},\ \bibinfo
  {pages} {075317} (\bibinfo {year} {2007})}\BibitemShut {NoStop}%
\bibitem [{\citenamefont {Jain}(1989)}]{composite_fermion}%
  \BibitemOpen
  \bibfield  {author} {\bibinfo {author} {\bibfnamefont {J.~K.}\ \bibnamefont
  {Jain}},\ }\href {\doibase 10.1103/PhysRevLett.63.199} {\bibfield  {journal}
  {\bibinfo  {journal} {Phys. Rev. Lett.}\ }\textbf {\bibinfo {volume} {63}},\
  \bibinfo {pages} {199} (\bibinfo {year} {1989})}\BibitemShut {NoStop}%
\bibitem [{\citenamefont {Haldane}(1983)}]{haldane_hierarchy}%
  \BibitemOpen
  \bibfield  {author} {\bibinfo {author} {\bibfnamefont {F.~D.~M.}\
  \bibnamefont {Haldane}},\ }\href {\doibase 10.1103/PhysRevLett.51.605}
  {\bibfield  {journal} {\bibinfo  {journal} {Physical Review Letters}\
  }\textbf {\bibinfo {volume} {51}},\ \bibinfo {pages} {605} (\bibinfo {year}
  {1983})}\BibitemShut {NoStop}%
\bibitem [{\citenamefont {Trugman}\ and\ \citenamefont
  {Kivelson}(1985)}]{trugman85}%
  \BibitemOpen
  \bibfield  {author} {\bibinfo {author} {\bibfnamefont {S.~A.}\ \bibnamefont
  {Trugman}}\ and\ \bibinfo {author} {\bibfnamefont {S.}~\bibnamefont
  {Kivelson}},\ }\href {\doibase 10.1103/PhysRevB.31.5280} {\bibfield
  {journal} {\bibinfo  {journal} {Physical Review B}\ }\textbf {\bibinfo
  {volume} {31}},\ \bibinfo {pages} {5280} (\bibinfo {year}
  {1985})}\BibitemShut {NoStop}%
\bibitem [{\citenamefont {Greiter}\ \emph {et~al.}(1991)\citenamefont
  {Greiter}, \citenamefont {Wen},\ and\ \citenamefont {Wilczek}}]{greiter91}%
  \BibitemOpen
  \bibfield  {author} {\bibinfo {author} {\bibfnamefont {M.}~\bibnamefont
  {Greiter}}, \bibinfo {author} {\bibfnamefont {X.-G.}\ \bibnamefont {Wen}}, \
  and\ \bibinfo {author} {\bibfnamefont {F.}~\bibnamefont {Wilczek}},\ }\href
  {\doibase 10.1103/PhysRevLett.66.3205} {\bibfield  {journal} {\bibinfo
  {journal} {Physical Review Letters}\ }\textbf {\bibinfo {volume} {66}},\
  \bibinfo {pages} {3205} (\bibinfo {year} {1991})}\BibitemShut {NoStop}%
\bibitem [{Note1()}]{Note1}%
  \BibitemOpen
  \bibinfo {note} {By this we mean that the ground state is exactly
  known.}\BibitemShut {Stop}%
\bibitem [{\citenamefont {MacDonald}(1994)}]{macdonald94}%
  \BibitemOpen
  \bibfield  {author} {\bibinfo {author} {\bibfnamefont {A.~H.}\ \bibnamefont
  {MacDonald}},\ }\href {http://arxiv.org/abs/cond-mat/9410047} {\bibfield
  {journal} {\bibinfo  {journal} {arXiv:cond-mat/9410047}\ } (\bibinfo {year}
  {1994})}\BibitemShut {NoStop}%
\bibitem [{\citenamefont {Rezayi}\ and\ \citenamefont
  {Haldane}(1994)}]{rezayi_haldane94}%
  \BibitemOpen
  \bibfield  {author} {\bibinfo {author} {\bibfnamefont {E.~H.}\ \bibnamefont
  {Rezayi}}\ and\ \bibinfo {author} {\bibfnamefont {F.~D.~M.}\ \bibnamefont
  {Haldane}},\ }\href {\doibase 10.1103/PhysRevB.50.17199} {\bibfield
  {journal} {\bibinfo  {journal} {Physical Review B}\ }\textbf {\bibinfo
  {volume} {50}},\ \bibinfo {pages} {17199} (\bibinfo {year}
  {1994})}\BibitemShut {NoStop}%
\bibitem [{\citenamefont {Lee}\ and\ \citenamefont
  {Leinaas}(2004)}]{Lee_Leinaas04}%
  \BibitemOpen
  \bibfield  {author} {\bibinfo {author} {\bibfnamefont {D.-H.}\ \bibnamefont
  {Lee}}\ and\ \bibinfo {author} {\bibfnamefont {J.~M.}\ \bibnamefont
  {Leinaas}},\ }\href {\doibase 10.1103/PhysRevLett.92.096401} {\bibfield
  {journal} {\bibinfo  {journal} {Phys. Rev. Lett.}\ }\textbf {\bibinfo
  {volume} {92}},\ \bibinfo {pages} {096401} (\bibinfo {year}
  {2004})}\BibitemShut {NoStop}%
\bibitem [{\citenamefont {Qi}(2011)}]{qi11}%
  \BibitemOpen
  \bibfield  {author} {\bibinfo {author} {\bibfnamefont {X.-L.}\ \bibnamefont
  {Qi}},\ }\href {\doibase 10.1103/PhysRevLett.107.126803} {\bibfield
  {journal} {\bibinfo  {journal} {Physical Review Letters}\ }\textbf {\bibinfo
  {volume} {107}},\ \bibinfo {pages} {126803} (\bibinfo {year}
  {2011})}\BibitemShut {NoStop}%
\bibitem [{\citenamefont {Wang}\ and\ \citenamefont
  {Scarola}(2011)}]{scarola11}%
  \BibitemOpen
  \bibfield  {author} {\bibinfo {author} {\bibfnamefont {H.}~\bibnamefont
  {Wang}}\ and\ \bibinfo {author} {\bibfnamefont {V.~W.}\ \bibnamefont
  {Scarola}},\ }\href {\doibase 10.1103/PhysRevB.83.245109} {\bibfield
  {journal} {\bibinfo  {journal} {Physical Review B}\ }\textbf {\bibinfo
  {volume} {83}},\ \bibinfo {pages} {245109} (\bibinfo {year}
  {2011})}\BibitemShut {NoStop}%
\bibitem [{\citenamefont {Seidel}\ \emph {et~al.}(2005)\citenamefont {Seidel},
  \citenamefont {Fu}, \citenamefont {Lee}, \citenamefont {Leinaas},\ and\
  \citenamefont {Moore}}]{seidel1}%
  \BibitemOpen
  \bibfield  {author} {\bibinfo {author} {\bibfnamefont {A.}~\bibnamefont
  {Seidel}}, \bibinfo {author} {\bibfnamefont {H.}~\bibnamefont {Fu}}, \bibinfo
  {author} {\bibfnamefont {D.-H.}\ \bibnamefont {Lee}}, \bibinfo {author}
  {\bibfnamefont {J.~M.}\ \bibnamefont {Leinaas}}, \ and\ \bibinfo {author}
  {\bibfnamefont {J.}~\bibnamefont {Moore}},\ }\href {\doibase
  10.1103/PhysRevLett.95.266405} {\bibfield  {journal} {\bibinfo  {journal}
  {Physical Review Letters}\ }\textbf {\bibinfo {volume} {95}},\ \bibinfo
  {pages} {266405} (\bibinfo {year} {2005})}\BibitemShut {NoStop}%
\bibitem [{\citenamefont {Haldane}(2011)}]{haldane11}%
  \BibitemOpen
  \bibfield  {author} {\bibinfo {author} {\bibfnamefont {F.~D.~M.}\
  \bibnamefont {Haldane}},\ }\href {\doibase 10.1103/PhysRevLett.107.116801}
  {\bibfield  {journal} {\bibinfo  {journal} {Physical Review Letters}\
  }\textbf {\bibinfo {volume} {107}},\ \bibinfo {pages} {116801} (\bibinfo
  {year} {2011})}\BibitemShut {NoStop}%
\bibitem [{\citenamefont {Qiu}\ \emph {et~al.}(2012)\citenamefont {Qiu},
  \citenamefont {Haldane}, \citenamefont {Wan}, \citenamefont {Yang},\ and\
  \citenamefont {Yi}}]{kunyang12}%
  \BibitemOpen
  \bibfield  {author} {\bibinfo {author} {\bibfnamefont {R.-Z.}\ \bibnamefont
  {Qiu}}, \bibinfo {author} {\bibfnamefont {F.~D.~M.}\ \bibnamefont {Haldane}},
  \bibinfo {author} {\bibfnamefont {X.}~\bibnamefont {Wan}}, \bibinfo {author}
  {\bibfnamefont {K.}~\bibnamefont {Yang}}, \ and\ \bibinfo {author}
  {\bibfnamefont {S.}~\bibnamefont {Yi}},\ }\href {\doibase
  10.1103/PhysRevB.85.115308} {\bibfield  {journal} {\bibinfo  {journal}
  {Physical Review B}\ }\textbf {\bibinfo {volume} {85}},\ \bibinfo {pages}
  {115308} (\bibinfo {year} {2012})}\BibitemShut {NoStop}%
\bibitem [{Note2()}]{Note2}%
  \BibitemOpen
  \bibinfo {note} {We note though a tractable truncated version of Eq.~(\ref
  {V1_cyl}) with matrix product ground state given in Ref. \protect
  \rev@citealpnum {Bergholtz12}.}\BibitemShut {Stop}%
\bibitem [{\citenamefont {Bernevig}\ and\ \citenamefont
  {Haldane}(2008{\natexlab{a}})}]{bernevig08}%
  \BibitemOpen
  \bibfield  {author} {\bibinfo {author} {\bibfnamefont {B.~A.}\ \bibnamefont
  {Bernevig}}\ and\ \bibinfo {author} {\bibfnamefont {F.~D.~M.}\ \bibnamefont
  {Haldane}},\ }\href {\doibase 10.1103/PhysRevLett.100.246802} {\bibfield
  {journal} {\bibinfo  {journal} {Physical Review Letters}\ }\textbf {\bibinfo
  {volume} {100}},\ \bibinfo {pages} {246802} (\bibinfo {year}
  {2008}{\natexlab{a}})}\BibitemShut {NoStop}%
\bibitem [{\citenamefont {Bernevig}\ and\ \citenamefont
  {Haldane}(2008{\natexlab{b}})}]{bernevig08-2}%
  \BibitemOpen
  \bibfield  {author} {\bibinfo {author} {\bibfnamefont {B.~A.}\ \bibnamefont
  {Bernevig}}\ and\ \bibinfo {author} {\bibfnamefont {F.~D.~M.}\ \bibnamefont
  {Haldane}},\ }\href {\doibase 10.1103/PhysRevB.77.184502} {\bibfield
  {journal} {\bibinfo  {journal} {Physical Review B}\ }\textbf {\bibinfo
  {volume} {77}},\ \bibinfo {pages} {184502} (\bibinfo {year}
  {2008}{\natexlab{b}})}\BibitemShut {NoStop}%
\bibitem [{\citenamefont {Haldane}\ and\ \citenamefont
  {Rezayi}(1985)}]{haldane85}%
  \BibitemOpen
  \bibfield  {author} {\bibinfo {author} {\bibfnamefont {F.~D.~M.}\
  \bibnamefont {Haldane}}\ and\ \bibinfo {author} {\bibfnamefont {E.~H.}\
  \bibnamefont {Rezayi}},\ }\href {\doibase 10.1103/PhysRevB.31.2529}
  {\bibfield  {journal} {\bibinfo  {journal} {Physical Review B}\ }\textbf
  {\bibinfo {volume} {31}},\ \bibinfo {pages} {2529} (\bibinfo {year}
  {1985})}\BibitemShut {NoStop}%
\bibitem [{\citenamefont {Wen}\ and\ \citenamefont {Niu}(1990)}]{wen_niu90}%
  \BibitemOpen
  \bibfield  {author} {\bibinfo {author} {\bibfnamefont {X.~G.}\ \bibnamefont
  {Wen}}\ and\ \bibinfo {author} {\bibfnamefont {Q.}~\bibnamefont {Niu}},\
  }\href {\doibase 10.1103/PhysRevB.41.9377} {\bibfield  {journal} {\bibinfo
  {journal} {Physical Review B}\ }\textbf {\bibinfo {volume} {41}},\ \bibinfo
  {pages} {9377} (\bibinfo {year} {1990})}\BibitemShut {NoStop}%
\bibitem [{\citenamefont {Read}\ and\ \citenamefont
  {Rezayi}(1996)}]{read_rezayi96}%
  \BibitemOpen
  \bibfield  {author} {\bibinfo {author} {\bibfnamefont {N.}~\bibnamefont
  {Read}}\ and\ \bibinfo {author} {\bibfnamefont {E.}~\bibnamefont {Rezayi}},\
  }\href {\doibase 10.1103/PhysRevB.54.16864} {\bibfield  {journal} {\bibinfo
  {journal} {Physical Review B}\ }\textbf {\bibinfo {volume} {54}},\ \bibinfo
  {pages} {16864} (\bibinfo {year} {1996})}\BibitemShut {NoStop}%
\bibitem [{\citenamefont {Bergholtz}\ and\ \citenamefont
  {Karlhede}(2005)}]{karlhede1}%
  \BibitemOpen
  \bibfield  {author} {\bibinfo {author} {\bibfnamefont {E.~J.}\ \bibnamefont
  {Bergholtz}}\ and\ \bibinfo {author} {\bibfnamefont {A.}~\bibnamefont
  {Karlhede}},\ }\href {http://link.aps.org/doi/10.1103/PhysRevLett.94.026802}
  {\bibfield  {journal} {\bibinfo  {journal} {Physcal Review Letters}\ }\textbf
  {\bibinfo {volume} {94}},\ \bibinfo {pages} {26802} (\bibinfo {year}
  {2005})}\BibitemShut {NoStop}%
\bibitem [{\citenamefont {Seidel}\ and\ \citenamefont
  {Lee}(2006)}]{seidel_dunghai06}%
  \BibitemOpen
  \bibfield  {author} {\bibinfo {author} {\bibfnamefont {A.}~\bibnamefont
  {Seidel}}\ and\ \bibinfo {author} {\bibfnamefont {D.-H.}\ \bibnamefont
  {Lee}},\ }\href {\doibase 10.1103/PhysRevLett.97.056804} {\bibfield
  {journal} {\bibinfo  {journal} {Physical Review Letters}\ }\textbf {\bibinfo
  {volume} {97}},\ \bibinfo {pages} {056804} (\bibinfo {year}
  {2006})}\BibitemShut {NoStop}%
\bibitem [{\citenamefont {Bergholtz}\ and\ \citenamefont
  {Karlhede}(2006)}]{karlhede2}%
  \BibitemOpen
  \bibfield  {author} {\bibinfo {author} {\bibfnamefont {E.~J.}\ \bibnamefont
  {Bergholtz}}\ and\ \bibinfo {author} {\bibfnamefont {A.}~\bibnamefont
  {Karlhede}},\ }\href {http://iopscience.iop.org/1742-5468/2006/04/L04001}
  {\bibfield  {journal} {\bibinfo  {journal} {J. Stat. Mech.}\ }\textbf
  {\bibinfo {volume} {L04001}} (\bibinfo {year} {2006})}\BibitemShut {NoStop}%
\bibitem [{\citenamefont {Seidel}\ and\ \citenamefont
  {Lee}(2007)}]{seidel_dunghai07}%
  \BibitemOpen
  \bibfield  {author} {\bibinfo {author} {\bibfnamefont {A.}~\bibnamefont
  {Seidel}}\ and\ \bibinfo {author} {\bibfnamefont {D.-H.}\ \bibnamefont
  {Lee}},\ }\href {\doibase 10.1103/PhysRevB.76.155101} {\bibfield  {journal}
  {\bibinfo  {journal} {Phys. Rev. B}\ }\textbf {\bibinfo {volume} {76}},\
  \bibinfo {pages} {155101} (\bibinfo {year} {2007})}\BibitemShut {NoStop}%
\bibitem [{\citenamefont {Seidel}(2008)}]{seidel2008a}%
  \BibitemOpen
  \bibfield  {author} {\bibinfo {author} {\bibfnamefont {A.}~\bibnamefont
  {Seidel}},\ }\href {http://link.aps.org/doi/10.1103/PhysRevLett.101.196802}
  {\bibfield  {journal} {\bibinfo  {journal} {Physical Review Letters}\
  }\textbf {\bibinfo {volume} {101}},\ \bibinfo {pages} {196802} (\bibinfo
  {year} {2008})}\BibitemShut {NoStop}%
\bibitem [{\citenamefont {Flavin}\ and\ \citenamefont
  {Seidel}(2011)}]{john_seidel11}%
  \BibitemOpen
  \bibfield  {author} {\bibinfo {author} {\bibfnamefont {J.}~\bibnamefont
  {Flavin}}\ and\ \bibinfo {author} {\bibfnamefont {A.}~\bibnamefont
  {Seidel}},\ }\href {\doibase 10.1103/PhysRevX.1.021015} {\bibfield  {journal}
  {\bibinfo  {journal} {Physical Review X}\ }\textbf {\bibinfo {volume} {1}},\
  \bibinfo {pages} {021015} (\bibinfo {year} {2011})}\BibitemShut {NoStop}%
\bibitem [{\citenamefont {Flavin}\ \emph {et~al.}(2012)\citenamefont {Flavin},
  \citenamefont {Thomale},\ and\ \citenamefont {Seidel}}]{john_seidel12}%
  \BibitemOpen
  \bibfield  {author} {\bibinfo {author} {\bibfnamefont {J.}~\bibnamefont
  {Flavin}}, \bibinfo {author} {\bibfnamefont {R.}~\bibnamefont {Thomale}}, \
  and\ \bibinfo {author} {\bibfnamefont {A.}~\bibnamefont {Seidel}},\ }\href
  {\doibase 10.1103/PhysRevB.86.125316} {\bibfield  {journal} {\bibinfo
  {journal} {Physical Review B}\ }\textbf {\bibinfo {volume} {86}},\ \bibinfo
  {pages} {125316} (\bibinfo {year} {2012})}\BibitemShut {NoStop}%
\bibitem [{\citenamefont {Seidel}\ and\ \citenamefont {Yang}(2011)}]{seidel11}%
  \BibitemOpen
  \bibfield  {author} {\bibinfo {author} {\bibfnamefont {A.}~\bibnamefont
  {Seidel}}\ and\ \bibinfo {author} {\bibfnamefont {K.}~\bibnamefont {Yang}},\
  }\href {\doibase 10.1103/PhysRevB.84.085122} {\bibfield  {journal} {\bibinfo
  {journal} {Physical Review B}\ }\textbf {\bibinfo {volume} {84}},\ \bibinfo
  {pages} {085122} (\bibinfo {year} {2011})}\BibitemShut {NoStop}%
\bibitem [{\citenamefont {Read}(2009)}]{read09}%
  \BibitemOpen
  \bibfield  {author} {\bibinfo {author} {\bibfnamefont {N.}~\bibnamefont
  {Read}},\ }\href {\doibase 10.1103/PhysRevB.79.045308} {\bibfield  {journal}
  {\bibinfo  {journal} {Physical Review B}\ }\textbf {\bibinfo {volume} {79}},\
  \bibinfo {pages} {045308} (\bibinfo {year} {2009})}\BibitemShut {NoStop}%
\bibitem [{Note3()}]{Note3}%
  \BibitemOpen
  \bibinfo {note} {Indeed, as formulated at present, these different Landau
  levels do not even live in the same Hilbert space, since the domain of the
  underlying wave functions depends on the value of $r$. This is
  inconsequential at present, however, and will later be remedied.}\BibitemShut
  {Stop}%
\bibitem [{\citenamefont {Read}\ and\ \citenamefont
  {Rezayi}(2011)}]{read_rezayi11}%
  \BibitemOpen
  \bibfield  {author} {\bibinfo {author} {\bibfnamefont {N.}~\bibnamefont
  {Read}}\ and\ \bibinfo {author} {\bibfnamefont {E.~H.}\ \bibnamefont
  {Rezayi}},\ }\href {\doibase 10.1103/PhysRevB.84.085316} {\bibfield
  {journal} {\bibinfo  {journal} {Physical Review B}\ }\textbf {\bibinfo
  {volume} {84}},\ \bibinfo {pages} {085316} (\bibinfo {year}
  {2011})}\BibitemShut {NoStop}%
\bibitem [{\citenamefont {Seidel}(2010)}]{seidel_duality}%
  \BibitemOpen
  \bibfield  {author} {\bibinfo {author} {\bibfnamefont {A.}~\bibnamefont
  {Seidel}},\ }\href {\doibase 10.1103/PhysRevLett.105.026802} {\bibfield
  {journal} {\bibinfo  {journal} {Physical Review Letters}\ }\textbf {\bibinfo
  {volume} {105}},\ \bibinfo {pages} {026802} (\bibinfo {year}
  {2010})}\BibitemShut {NoStop}%
\bibitem [{Note4()}]{Note4}%
  \BibitemOpen
  \bibinfo {note} {It turns out that the final form of $\protect \mathaccentV
  {tilde}07E{{\protect \bf G}}_\tau $ also contains a one body part that we
  omit in \protect \textup {\hbox {\mathsurround \z@ \protect \normalfont
  (\ignorespaces \ref {Gmmnn}\unskip \@@italiccorr )}}, \protect \textup {\hbox
  {\mathsurround \z@ \protect \normalfont (\ignorespaces \ref {Gtau1}\unskip
  \@@italiccorr )}} for brevity. However this part transforms
  analogously.}\BibitemShut {Stop}%
\bibitem [{Note5()}]{Note5}%
  \BibitemOpen
  \bibinfo {note} {We are indebted to G. M{\"o}ller for this
  observation.}\BibitemShut {Stop}%
\bibitem [{\citenamefont {Haldane}(2009)}]{haldane09}%
  \BibitemOpen
  \bibfield  {author} {\bibinfo {author} {\bibfnamefont {F.~D.~M.}\
  \bibnamefont {Haldane}},\ }\href {http://arxiv.org/abs/0906.1854} {\bibfield
  {journal} {\bibinfo  {journal} {arXiv:0906.1854}\ } (\bibinfo {year}
  {2009})}\BibitemShut {NoStop}%
\bibitem [{\citenamefont {Avron}\ \emph {et~al.}(1995)\citenamefont {Avron},
  \citenamefont {Seiler},\ and\ \citenamefont {Zograf}}]{avron95}%
  \BibitemOpen
  \bibfield  {author} {\bibinfo {author} {\bibfnamefont {J.~E.}\ \bibnamefont
  {Avron}}, \bibinfo {author} {\bibfnamefont {R.}~\bibnamefont {Seiler}}, \
  and\ \bibinfo {author} {\bibfnamefont {P.~G.}\ \bibnamefont {Zograf}},\
  }\href {\doibase 10.1103/PhysRevLett.75.697} {\bibfield  {journal} {\bibinfo
  {journal} {Physical Review Letters}\ }\textbf {\bibinfo {volume} {75}},\
  \bibinfo {pages} {697} (\bibinfo {year} {1995})}\BibitemShut {NoStop}%
\bibitem [{\citenamefont {Bradlyn}\ \emph {et~al.}(2012)\citenamefont
  {Bradlyn}, \citenamefont {Goldstein},\ and\ \citenamefont {Read}}]{read12}%
  \BibitemOpen
  \bibfield  {author} {\bibinfo {author} {\bibfnamefont {B.}~\bibnamefont
  {Bradlyn}}, \bibinfo {author} {\bibfnamefont {M.}~\bibnamefont {Goldstein}},
  \ and\ \bibinfo {author} {\bibfnamefont {N.}~\bibnamefont {Read}},\ }\href
  {\doibase 10.1103/PhysRevB.86.245309} {\bibfield  {journal} {\bibinfo
  {journal} {Physical Review B}\ }\textbf {\bibinfo {volume} {86}},\ \bibinfo
  {pages} {245309} (\bibinfo {year} {2012})}\BibitemShut {NoStop}%
\bibitem [{\citenamefont {L{\'e}vay}(1995)}]{levay95}%
  \BibitemOpen
  \bibfield  {author} {\bibinfo {author} {\bibfnamefont {P.}~\bibnamefont
  {L{\'e}vay}},\ }\href {\doibase doi:10.1063/1.531066} {\bibfield  {journal}
  {\bibinfo  {journal} {Journal of Mathematical Physics}\ }\textbf {\bibinfo
  {volume} {36}},\ \bibinfo {pages} {2792} (\bibinfo {year}
  {1995})}\BibitemShut {NoStop}%
\bibitem [{Note6()}]{Note6}%
  \BibitemOpen
  \bibinfo {note} {The latter case is obviously realized for the Laughlin state
  on the cylinder and the operator $G_{r^{-2}}$ defined in Sec. \ref
  {cylinder}, where the expression analogous to the right hand side Eq.~(\ref
  {presentation2}) leaves the thin torus limiting state invariant.}\BibitemShut
  {Stop}%
\bibitem [{\citenamefont {Bernard}(1988)}]{bernard88}%
  \BibitemOpen
  \bibfield  {author} {\bibinfo {author} {\bibfnamefont {D.}~\bibnamefont
  {Bernard}},\ }\href {\doibase 10.1016/0550-3213(88)90217-9} {\bibfield
  {journal} {\bibinfo  {journal} {Nuclear Physics B}\ }\textbf {\bibinfo
  {volume} {303}},\ \bibinfo {pages} {77} (\bibinfo {year} {1988})}\BibitemShut
  {NoStop}%
\bibitem [{\citenamefont {Nakamura}\ \emph {et~al.}(2012)\citenamefont
  {Nakamura}, \citenamefont {Wang},\ and\ \citenamefont
  {Bergholtz}}]{Bergholtz12}%
  \BibitemOpen
  \bibfield  {author} {\bibinfo {author} {\bibfnamefont {M.}~\bibnamefont
  {Nakamura}}, \bibinfo {author} {\bibfnamefont {Z.-Y.}\ \bibnamefont {Wang}},
  \ and\ \bibinfo {author} {\bibfnamefont {E.~J.}\ \bibnamefont {Bergholtz}},\
  }\href {\doibase 10.1103/PhysRevLett.109.016401} {\bibfield  {journal}
  {\bibinfo  {journal} {Physical Review Letters}\ }\textbf {\bibinfo {volume}
  {109}},\ \bibinfo {pages} {016401} (\bibinfo {year} {2012})}\BibitemShut
  {NoStop}%
\end{thebibliography}%
\end{document}